\documentclass[12pt,a4paper]{article}

\usepackage[left=2.4cm,right=2.4cm,top=3cm,bottom=2.5cm]{geometry}
\usepackage[font=small,labelfont=bf]{caption}
\usepackage{framed,multirow,bigstrut,makecell} % graphicx,
\usepackage{diagbox}
\usepackage{latexsym}
\usepackage{url}
\usepackage{xcolor}
\usepackage{graphicx}
\graphicspath{{Images/}}
\usepackage[affil-it]{authblk}

\usepackage{amssymb, amsfonts, stmaryrd, fancyhdr, amsmath, amsthm, amsbsy, amsfonts, titletoc} 
\allowdisplaybreaks[4]
\newtheorem{theorem}{Theorem}[section]

\theoremstyle{definition}

\newtheorem{remark}{Remark}[section]

\numberwithin{equation}{section}

\usepackage{sectsty}
\sectionfont{\fontsize{15}{15}\selectfont}
\subsectionfont{\fontsize{13}{15}\selectfont}

\def\bphi{\overline{\phi}} 
\def\ep{\varepsilon}

\def\nP{n_P}

\def\NA{{N_A}}
\def\NB{{N_B}}
\def\nS{n_S}
\def\mG{\Delta}
\def\mL{\mathcal{L}}
\def\cN{\mathcal{N}}
\def\mR{\mathcal{R}}
\def\mE{\mathcal E_{\textrm{FCH}}}
\def\cE{\mathcal E}
\def\Wq{W_{\textrm q}}
\def\Wscmf{W_{\textrm{S}}}

\def\cF{{\mathcal F}}
\def\IMEX{\textrm IMEX }
\def\PSD{\textrm PSD }
\def\SAV{\textrm SAV }
\def\ETD{\textrm ETD }
\def\N{\mathcal{N}}
\def\md{\text{d}}
\def\me{\text{e}}

\newcommand{\dcoef}{\textrm{d}}
\newcommand{\qtype}{\textrm{q}}

\newcommand\keywords[1]{~~~\begin{minipage}{0.87\textwidth}{\bf Key words}: #1 \end{minipage}}

\title{\bf \fontsize{16}{15}\selectfont Benchmark Computation of Morphological Complexity in the Functionalized Cahn-Hilliard Gradient Flow}

\author[1,2]{\fontsize{14}{15}\selectfont Andrew Christlieb}
\author[1]{Keith Promislow}
\author[3]{Zengqiang Tan}
\author[4,5]{Sulin Wang
\thanks{Corresponding author. \\ 
Emails: 
{\tt christli@msu.edu}(A. Christlieb), 
{\tt promislo@msu.edu} (K. Promislow), 
{\tt tzengqiang@163.com} (Z. Tan), 
{\tt sulinw@mtu.edu} (S. Wang), 
{\tt wetton@math.ubc.ca} (B. Wetton), 
{\tt swise1@utk.edu} (S.M. Wise).}}
\author[6]{Brian Wetton}
\author[7]{Steven M. Wise}

\affil[1]{\normalsize Department of Mathematics, Michigan State University, East Lansing, MI 48824, USA.}
\affil[2]{Department of Computational Mathematics, Science and Engineering, Michigan State University, East Lansing, MI 48824, USA.}
\affil[3]{School of Science, Wuhan University of Technology, Wuhan 430070, China.}
\affil[4]{School of Mathematics, Hunan University, Changsha 410082, China.}
\affil[5]{Hunan Provincial Key Laboratory of Intelligent Information Processing and Applied Mathematics, Hunan University, Changsha 410082, China.}
\affil[6]{Department of Mathematics, The University of British Columbia, Vancouver, BC V6T 1Z2, Canada.}
\affil[7]{Department of Mathematics, The University of Tennessee, Knoxville, TN 37996, USA.}
\date{}

\usepackage{setspace}
\makeatletter
\let\oldaffillist\AB@affillist
\renewcommand{\AB@affillist}{
\begin{flushleft}
	\begin{spacing}{1.1}
	\oldaffillist 
	\end{spacing}
\end{flushleft}}
\makeatother

\begin{document}
\maketitle
\vspace{-1.5cm}

%%%%% Begin Abstract %%%%%%%%%%%
\begin{abstract}
Reductions of the self-consistent mean field theory model of amphiphilic molecules in solvent can lead to a singular family of functionalized Cahn-Hilliard (FCH) energies. We modify these energies, mollifying the singularities to stabilize the computation of the gradient flows and develop a series of benchmark problems that emulate the ``morphological complexity'' observed in experiments. % reported in \cite{Bates-BD}. 
These benchmarks investigate the delicate balance between the rate of absorption of amphiphilic material onto an interface and a least energy mechanism to disperse the arriving mass. The result is a trichotomy of responses in which two-dimensional interfaces either lengthen by a regularized motion against curvature, undergo pearling bifurcations, or split directly into networks of interfaces. We evaluate a number of schemes that use second order backward differentiation formula (BDF2) type time stepping coupled with Fourier pseudo-spectral spatial discretization. The BDF2-type schemes are either based on a fully implicit time discretization with a preconditioned steepest descent (\PSD\!\!) nonlinear solver or upon linearly implicit time discretization based on the standard implicit-explicit (\IMEX\!\!) and the scalar auxiliary variable (\SAV\!\!) approaches. We add an exponential time differencing (ETD) scheme for comparison purposes.  All schemes use a fixed local truncation error target with adaptive time-stepping to achieve the error target. Each scheme requires proper ``preconditioning'' to achieve robust performance that can enhance efficiency by several orders of magnitude. The nonlinear \PSD scheme achieves the smallest global discretization error at fixed local truncation error, however the \IMEX and \SAV schemes are the most computationally efficient as measured by the number of Fast Fourier Transform (FFT) calls required to achieve a desired global error. Indeed the performance of the \SAV scheme directly mirrors that of \IMEX\!\!, modulo a factor of 1.4 in FFT calls for the auxiliary variable system.
\end{abstract}
%%%%% end %%%%%%%%%%%

%%%%% AMS/PACs/Keywords %%%%%%%%%%%
%\pac{}
%\ams{35K35, 65M06, 65M12, 65M50}
\keywords{phase field model,
benchmark computations,
adaptive time stepping,
functionalized Cahn-Hilliard.}
%%%% maketitle %%%%%
\maketitle

%%%% Start %%%%%%
\section{Introduction}
We present a series of physically motivated computational benchmark  problems addressing the evolution of the functionalized Cahn-Hilliard (FCH) gradient flow. This system supports  families of equilibria with rich morphological structure separated by slightly different energies. The faithful resolution of final end states requires significant computational accuracy. There has been considerable recent attention to the development of energy stable computational schemes for gradient descent flows \cite{DJLQ-18, FGLWWC, GZW-p, SXU-18, SX-18, V-LR03, W-10, ZOWW}. 
Gradient flows are defined by the dissipation of a free energy, and it is essential that numerical schemes preserve that property. Energy stable schemes have the desirable property that the energy, or a modified energy, decreases at every time-step irrespective of time-step size. 
We argue that where possible energy decay should be a consequence of accuracy. In some situations energy decay without accuracy can lead to plausible but incorrect computational outcomes. 
Conversely accuracy should be balanced against computational cost. This motivates a comparison of computational efficiency between schemes as measured by the minimal computational cost required to achieve a desired global discretization error.

\begin{figure}[ht!]
\centering
\begin{tabular}{cp{2.5in}}
\includegraphics[width=3in]{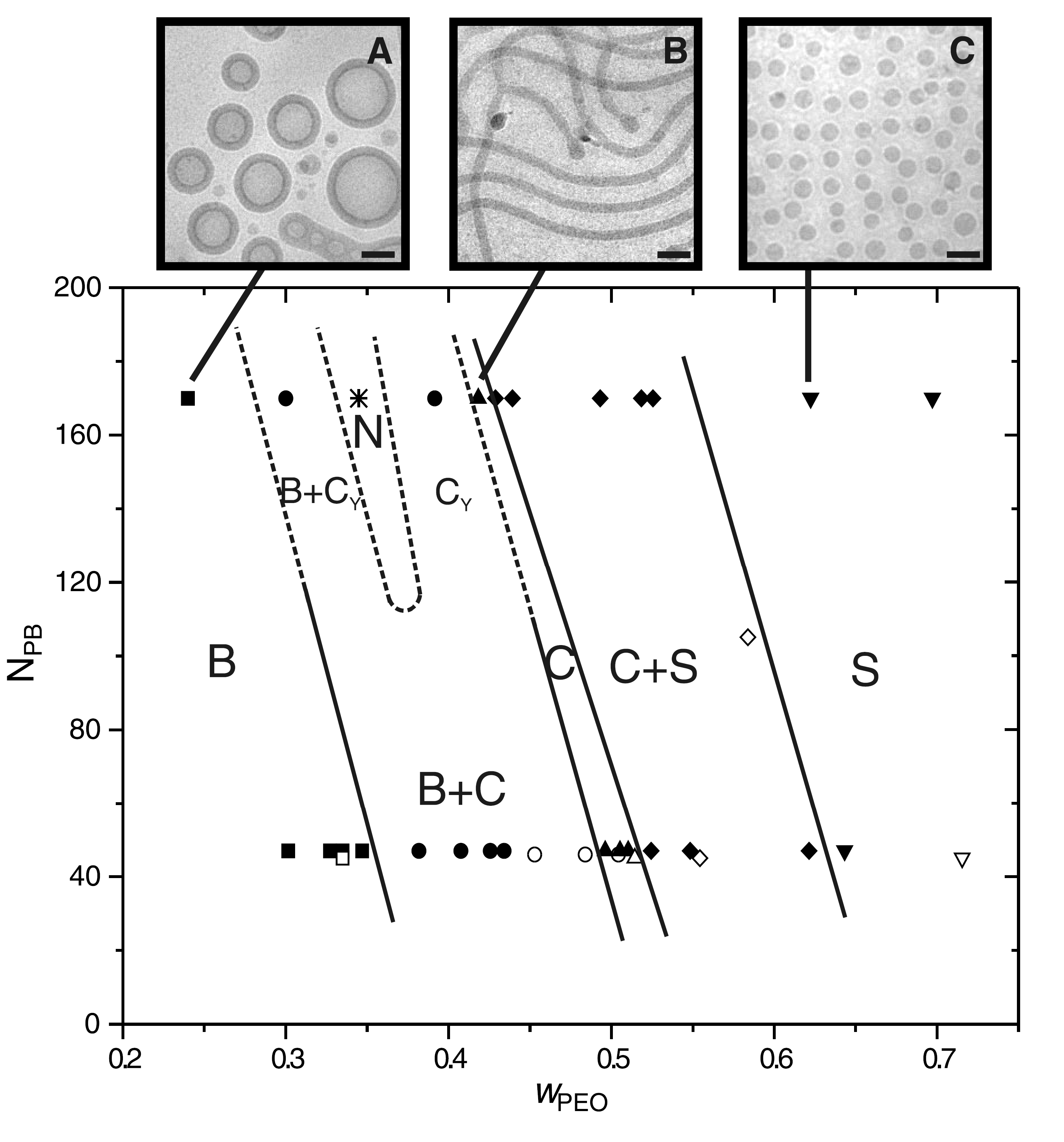} & \\
\end{tabular}
\vskip -3.3in
\begin{tabular}{p{3.0in}p{2.1in}}
&\includegraphics[width=2in,height=1.5in]{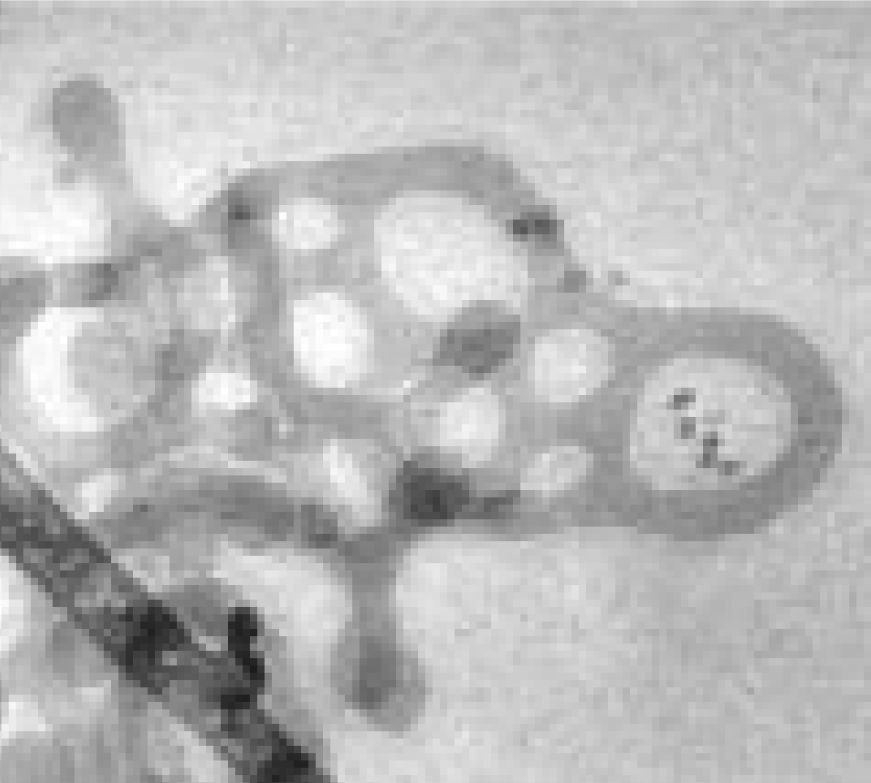} \\
&\includegraphics[width=2in,height=1.5in]{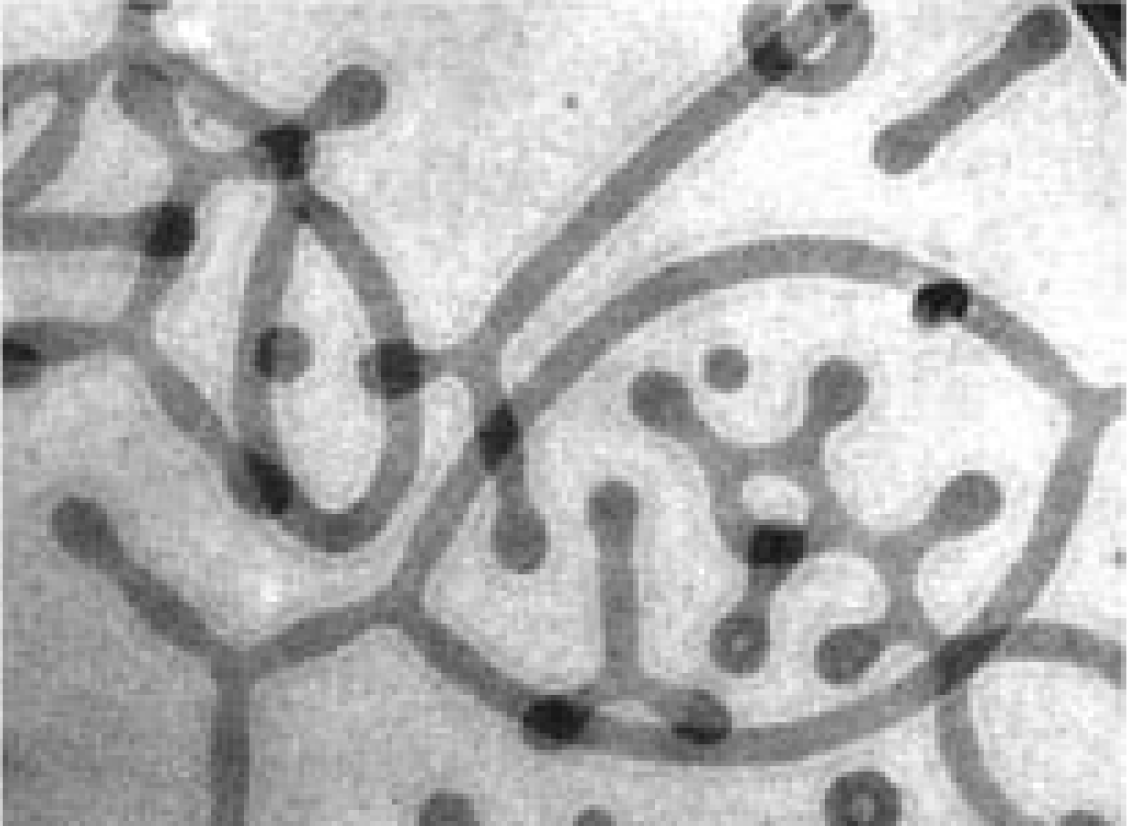} 
\end{tabular}
\vskip 0in
\caption{\small 
{\bf (left)} Experimentally observed bifurcation diagram for the morphology of blends of Polyethylene oxide (PEO) - Polybutadiene (PB) amphiphilic diblock in water.  The horizontal axis, $w_{\rm PEO}$, is the weight fraction of PEO as a percent of the total diblock weight, and the vertical axis denotes the molecular weights of the PB component of the diblock, fixed at $N_{\rm PB}=45$ or $170$ (vertical axis). 
\emph{Morphological Complexity} is observed for $N_{\rm PB}=170$ but not for the shorter $N_{\rm PB}=45$ chains.
{\bf (right)}  Experimental images from the morphological complexity regime showing (top) network structures and (bottom) a mixture of end caps and $Y$-junction morphology corresponding to regions marked $N$ and $C_Y$ in the bifurcation diagram. From Figures 1 and 2AC of \cite{Bates-BD}, Reprinted with permission from AAAS.}
%\vspace{0.2in}
\normalsize
\label{f:Bates}
\end{figure}

Meaningful assessment of computational efficiency can be achieved from gradient flows that harbour strong nonlinear interactions that generate selection mechanisms between distinct outcomes with small energy differences.  For motivation, we emulate the ``morphological complexity'' experiments presented in \cite{Bates-BD}. 
By strongly dispersing (stirring) amphiphilic diblock polymers in solvent, and then allowing the mixture to relax, the authors of that study observed the formation of a wide variety of structures whose evolution and end-state depend sensitively upon the polymer chain and mixture properties, see Figure\,\ref{f:Bates} and \cite{BD-Barnhill, Blanazs-09}. 
Reductions of the self-consistent mean field theory  models of amphiphilic molecules in solvent can lead to a singular family of FCH energies, \cite{CP-22}. We modify these energies, mollifying the singularities to produce a family of computationally tractable, but highly nonlinear, FCH gradient flows similar to those studied earlier, \cite{Dai-13, Dai-15, Gavish-11}. 
We present a series of benchmark problems that recover the onset of morphological complexity. These benchmarks are conducted in a regime in which interfacial width, controlled by $\varepsilon$, is small. They  reveal a delicate balance between the rate of absorption of amphiphilic material onto an interface and the gradient flow's selection of a least energy mechanism to redistribute the amphiphilic mass along the interface after absorption. This rate-based selection mechanism yields a trichotomy of responses in which two-dimensional interfaces either grow by a regularized motion against curvature, under-go pearling bifurcations the form structure within the, or directly curve-split into networks of interfaces. We present four numerical schemes, each combining second-order temporal discretization and pseudo-spectral spatial discretization. 
The FCH energy is computationally stiff due to the strength of its nonlinear terms. Each of the second order methods considered balance implicit and explicit terms. Their efficiency is sensitive to the choice of the implicit terms, with improvements of several orders of magnitude possible when the methods are well balanced. These methods include an implicit-explicit (BDF2-\IMEX\!\!) method, a second order exponential time differencing Runge-Kutta method (ETDRK2), and a scalar auxiliary variable approach (BDF2-\SAV\!\!). The latter scheme features provably unconditional modified energy stability properties. All of these schemes are linear in their implicit stage. We compare these with a fully implicit, second order, backward differentiation scheme based upon a preconditioned steepest descent with approximate line search (BDF2-\PSD\!\!) for the nonlinear solve. For brevity we drop the `BDF2' and `RK2' components of the acronyms in the sequel.

The FCH gradient flows possess distinct, emergent timescales that render fixed time-stepping approaches inefficient. For each scheme a specified target local truncation error is used to generate an adaptive time-stepping procedure. The first set of benchmarks, the sub-critical, critical, and super-critical, use relatively smooth potentials in the FCH energy, and vary the mass of amphiphilic material distributed within the background of the initial data. This serves to vary the rate of absorption of mass onto the interface. The supercritical benchmark has an absorption rate sufficient to trigger the defect-inducing bifurcations that are the genesis of morphological complexity. 
A proper resolution of the time evolution requires considerable accuracy. The second set of benchmarks enhances the stiffness of the FCH energy by increasing the convexity of the potential well at the background state, mimicking the singular nature of the FCH energy as reduced from the self-consistent mean field theory. This adds a small ``foot'' to the left minima of the well, see Figure\,\ref{f:W-qtype}, hence these benchmarks 
are called Foot 1 and Foot 2. The stiffness increases the ratio of the absorption rate to the mass redistribution rate affording a second mechanism to induce morphological complexity.

Each of the second order schemes we consider requires an appropriate choice of implicit terms or preconditioner. This choice is typically based upon the linearization about a spatially constant equilibrium solution. The linearly implicit \IMEX and \SAV accommodate the increase in stiffness for the Foot 1 and Foot 2 benchmarks without significant adjustment.  The nonlinear solve in the \PSD scheme requires optimization of internal parameters, in particular an error tolerance associated to the iterative nonlinear solver, to converge. Moreover the efficiency of the \PSD scheme decrease in comparison to the two linear implicit schemes with increasing numerical stiffness. Other preconditioning schemes, for example based upon non-constant coefficient linear terms, could improve the efficiency of the \PSD scheme, however this is not considered here. The \ETD approach was relatively insensitive to choice of implicit terms and less efficient at handling the nonlinear stiffness in the super-critical benchmark. It was not pursued for the Foot 1 and Foot 2 benchmarks.

We conduct grid refinement studies to verify that each benchmark has an adequate spatial resolution and develop highly accurate solutions for each benchmark by an extensive computation with a very small local truncation error. Once spatially resolved, all four schemes yield concordant results for sufficiently small specified local truncation error.  We adjust the local truncation error restriction and use short runs to tune performance parameters in each scheme for each benchmark, and record the accuracy and cost of each optimized scheme.  
At given local truncation error we find that the \PSD approach is generically the most accurate with \IMEX and \SAV generally the least accurate, as measured by global error at the final time. However, at fixed local truncation error the \IMEX and \SAV schemes require less computational effort than the \PSD and \ETD\!, with the \IMEX and \SAV schemes performing almost identically, modulo a fixed factor in extra computational effort required by \SAV due to the extra system for the auxiliary variable. 
For these benchmarks a global $L^2$ relative discretization error of $2.5\times10^{-3}$ is found to be a harbinger of global accuracy, and within this constraint we view the local truncation error as an internal parameter to be adapted for each scheme to optimize global performance. 
For the sub-critical, critical, and super-critical benchmarks, all schemes except \ETD achieved this global accuracy with comparable efficiency although at quite different values of the local truncation error. While it displays second order accuracy, the \ETD scheme does not seem to be competitive.  We present a heuristic argument in Appendix B that indicates that \ETD is more sensitive to the interface width parameter $\varepsilon$ in the thin interface limit $\varepsilon\ll1$ in which we compute.  As presented in Figure\,\ref{f:BM3-L2}, achieving this accuracy for the super-critical benchmark requires $1.5\times10^{5}, 2\times 10^{5}$, and $2.1\times10^5$ FFT calls for \IMEX, \PSD and \SAV respectively, while \ETD requires $2.5\times 10^6$ FFT calls. As the global error target is further tightened, the \PSD scheme requires increased computational effort, first increasing rapidly and then saturating. 
Conversely the computational effort of the \IMEX and \SAV schemes increases linearly with global discretization error. 
%related by a fixed factor of two.  
For the more strongly nonlinear Foot 1 and Foot 2 benchmarks the efficiency of the linear-implicit schemes continues its linear relationship to global discretization error. 
%and to each-other by the same factor of two. 
As depicted in Figure\,\ref{f:BM4-L2}, for the stronger nonlinearity the efficiency of \PSD deteriorated in comparison to the linear-implicit methods.

The \SAV scheme is specifically designed to be energy stable with respect to an associated modified energy. This property either assumes fixed time-stepping, which is impractical for the FCH gradient flows in cases for which accuracy is paramount, or an adaptive time stepping based upon modifications by factors of two and nesting. This latter strategy is implemented for the super-critical benchmark within the BDF2-SAV scheme. This was found to provide no benefit for accuracy while increasing computational cost by a factor of two to three. We also implement the second order Crank-Nicolson approach in combination with the SAV strategy but find that it is not computationally efficient. In all cases all convergent schemes preserve the energy decay property of the gradient flow.

\begin{remark}
The work \cite{ZOWW} directly compares the \PSD and \SAV methods described herein, but in the context of uniform, fixed time step setting. Based upon their experience with FCH-type simulations, the authors state that ``ultimately adaptive time stepping algorithms should be compared."  The present study seeks to fill this gap, using time step adaptivity to make quantitative comparison of accuracy against efficiency for a variety of numerical schemes.  Moreover, the family of regularized FCH models presented here allow for interpolation between the smooth versions of the FCH considered in earlier analytical and numerical studies and the singular versions arising as reductions from self-consistent mean field analysis whose inherent numerical stiffness makes them more challenging than the models considered in~\cite{ZOWW}.
\end{remark}

This paper is organized as follows. In section 2, we briefly sketch the derivation of a singular FCH model from a random phase approximation of self-consistent mean field theory, outline the regularization of the singular model and its use to calibrate the family of regularized FCH models studied herein. We also present the initial data and motivate the benchmark problems. This derivation illuminates the incorporation of the well-stiffness in the Foot 1 and Foot 2 benchmarks that is the initial motivation for this computational study.  In section 3, we present the second order adaptive numerical schemes that we use to resolve the benchmark problems and highlight the sensitivity of efficiency to choice of implicit terms. In section 4, we present an overview of the simulations of each of the five benchmark problems for a fixed local truncation error, showing the conditions under which the schemes agree and disagree. In section 5, we contrast the performance of the schemes, particularly with respect to accuracy in the far-field of the domain, energy decay, evaluation of the precise critical value for
onset of defects, and comparison of time-stepping performance and computational efficiency.  We summarize the performance in section 6. The appendixes provide proof of energy stability for the \SAV scheme and a heuristic analysis of time-stepping for \ETD and \IMEX in the thin interface regime $\varepsilon\ll1$.

\section{Mean field approximation of amphiphilic diblock suspensions}
The self consistent mean-field (SCMF) approach derives density functional models that approximate the bulk interactions of collections of polymers represented by molecular units, \cite{Fred-SCMFT}. When applied to amphiphilic diblock polymers suspended in a solvent the reduction yields a free energy for the three density components, $\phi_i$, for $i=A,B,S$, which represent the hydrophilic head, $A$, and the hydrophobic tail, $B$, of the diblock polymer, and the solvent, $S$, respectively. 
Considering a suspension of $n_s$ solvent molecules and $n_P$ polymer diblocks, each comprised of $N_A$ and $N_B$ monomers of 
molecule $A$ and $B$, respectively, \cite{CR-03, UD05a} used the self-consistent mean field reduction to derive the free energy to a continuum phase-field model.   More specifically,  they introduced the mean densities
\begin{equation}
\bphi_A=\frac{\nP\NA}{|\Omega|}, \quad \bphi_B=\frac{\nP\NB}{|\Omega|}, \quad  \bphi_S=\frac{\nS}{|\Omega|},
\end{equation}
and derived a bilinear approximation to the SCMF free energy expressed in terms of the variance from the mean  $\phi_{i0}=\phi_i-\bphi_i$,
\begin{equation}
\label{eq:SCMFred}
\cF^{(2)}_{\rm UD}(\phi_0) = \sum_{ij}\int_\Omega \frac{a_{ij}}{\sqrt{\bphi_i\bphi_j}} (D^{-1} \phi_{i0}) (D^{-1} \phi_{j0}) + \Big(\frac{b_{ij}}{\sqrt{\bphi_i\bphi_j}}+\chi_{ij}\Big)\phi_{i0}\phi_{j0} +\delta_{ij} \frac{c_{ij}}{\bphi_i} |\nabla \phi_{i0}|^2\, \text{d}x.
\end{equation}
Here $a=(a_{ij})$, $b=(b_{ij})$, $c=(c_{ij})$, with $i,j\in\{A, B, S\}$, denote material parameters and $\delta_{ij}$ is the usual Kronecker delta function. Their derivation
is similar to \cite{CR-05}, with both approaches incorporating  long-range interaction terms through the operator $D:=(-\Delta)^\frac12$, 
the square-root of the negative Laplacian operator, subject to periodic boundary conditions. The long-range terms describe entropic effects of chain folding and 
volume exclusion derived from the interactions of the polymer chains with effective mean fields. A similar energy was proposed as a model of a
microemulsions of oil, water, and surfactant by \cite{TS-87}, who argued directly, and somewhat phenomenologically, from a Landau theory for a scalar
density.  This bilinear model was extended to a nonlinear one by \cite{GG-94} and \cite{GS-90},  who proposed a density dependence on the coefficients. Uneyama and Doi
also proposed a nonlinear extension, \cite{UD05b}, for their vector model in which the average density $\bphi_k$ was replaced with the local density $\phi_k$. 
This extrapolation yields  a family of models that include the Ohta-Kawasaki free energies. A general description of this extrapolation is presented in \cite{Fred-SM22}. In \cite{CP-22} the nonlinear extrapolation approach was modified, first through a shift in dependent variables to the spatially averaged density 
$\psi_k:= D^{-1}\phi_{k0},$ 
and then by an extrapolation step in which the average density $\bphi_k$ is replaced with the slowly varying average density,
\begin{equation}
\bphi_k \rightarrow \bphi_k(1+\psi_k).
\end{equation}

The three-component model is then reduced to a scalar field similar to \cite{GS-90} by requiring a point-wise incompressibility, 
$\psi_A+\psi_B+\psi_S=0$, and replacing the global constraint on the $A$- and $B$-polymer fractions with the point-wise constraint,  $\phi_A/N_A=\phi_B/N_B$.
Choosing the parameterization
$$ \psi_A=\psi_B=\frac{(b_r-b_l)u+(b_r+b_l)}{2m_f},~ \psi_S=1-\frac{(b_r-b_l)u+(b_r+b_l)}{2},$$
in terms of the free variable $u$, for choices of $b_r>b_l$ made below that normalize the range of $u$.  The resulting model depends upon $N_P:=N_A+N_B$, the polymer fractions $\alpha_A=N_A/N_P$ and $\alpha_B=1-\alpha_A$, 
the polymer-solvent molecular mole fraction $m_f:=n_PN_P/n_S$, and the dimensionless parameter $\ep=\frac{l}{L}N_P^{1/2}\ll 1$  which rescales the Kuhn length $l$ of the diblock polymer into a mean-square end-to-end length a single ideal diblock polymer chain expressed as a ratio of the domain length $L$. 
The amphiphilicity of the diblock molecules is expressed in terms of a weighted Flory-Huggins parameter
\begin{equation}
\chi_w:=\alpha_A\chi_{AS} + \alpha_B\chi_{BS} - \alpha_A\alpha_B\chi_{AB}>0,
\end{equation}
where for $k,m\in\{A, B, S\}$ the Flory-Huggins parameters $\chi_{km}>0$ record the strength of the repulsive interaction between 
a $k$-monomer and an $m$-monomer.  The value of $\chi_w$ depends upon the composition of the polymer diblock chain, but not on its length.

With these reductions and notation, the Uneyama-Doi bilinear energy \eqref{eq:SCMFred} reduces to the singular functionalized Cahn-Hilliard (S-FCH) form
\begin{equation}
\mathcal F_{\mathrm{S-FCH}}(u)= \frac12 \int_\Omega  \left(\ep^2\Delta u - \Wscmf'(u)\right)^2  + P(u) \text{d}x,
\end{equation}
where the singular potential $\Wscmf$ is defined via its derivative,
\begin{equation}
\begin{aligned}
\Wscmf'(u) = m_f\Big [24 & \ln\big|(b_r-b_l)u+(b_r+b_l)+2m_f\big| \\
&- 6N_P\Big( \ln \big |(b_r-b_l)u+(b_r+b_l)-2 \big |+\chi_w (b_r-b_l) u\Big)\Big ]+C_0.
\end{aligned}
\end{equation}

The condition $\chi_w>0$  guarantees that $\Wscmf'$ has three zeros on its domain. The parameters $b_r$ and $b_l$ are chosen to map the left and right zeros to $-1$ and $+1$ respectively, and the potential $\Wscmf$ is defined as the primitive of $\Wscmf'$ that has a double zero at $u=-1$. 
The first derivatives of the well $\Wscmf$  are singular at the endpoints where the  corresponding to pure solvent and pure polymer phases.  The perturbative potential $P$ takes the form
\begin{equation}
P(u):= \frac{9(b_r-b_l)}{\alpha_A\alpha_B}\frac{u^2}{u(b_r-b_l)+2m_f}-\big (\Wscmf'(u)\big )^2.
\end{equation}
The constant $C_0$ does not impact the value of the energy and is chosen to minimize the perturbative potential $P$.

\subsection{Regularized FCH and experimental motivation for the benchmark problems}
We draw motivation for the benchmark simulations from the complexity observed in the experiments conducted in \cite{Bates-BD}.  
In that study the authors prepared well-stirred dispersions of amphiphilic diblock of Polyethylene oxide (PEO) - Polybutadiene (PB) in water, and
allowed the mixture to relax and come to quasi-equilibrium.  The weight fraction of polymer was fixed at 1\%, and they considered a long and a short polymer chain, 
characterized by a fixed molecular length of the hydrophobic PB, with $N_{\rm PB} (=N_B)$ taken as 45 and 170.  They varied the aspect ratio $\alpha_A=N_A/N_B$, 
characterized by the weight fraction, $w_{\text{PEO}}$, of the amphiphilic PEO component. They recovered a bifurcation diagram, presented in Figure\,\ref{f:Bates}
(left), which shows that for the short chains the well-mixed dispersions largely formed codimension one spherical bilayer interfaces, codimension two solid tubes, or codimension three solid spherical micelles, with some overlap depending upon the aspect ratio. However for $\alpha_A\in(0.3,0.5)$ the suspensions of
long chains form structures that are loaded with defects, such as the network structures and endcaps depicted in Figure\,\ref{f:Bates} (right - top and bottom).

The self-assembly of spatially extended morphologies from a relatively dilute suspension can be viewed as an absorption and a redistribution process. The dispersed amphiphilic molecules
are generically too dilute to self assemble, but may diffuse until they arrive at localized structure where they insert themselves to lower their contribution to the system energy by isolating their hydrophobic tail from contact with the solvent. Within the FCH model, the rate of absorption of mass onto the interface determines the final outcome of this growth phase. The selection mechanism for the end state is delicate, with many possible outcomes separated by slightly different final energies. This landscape affords an excellent diagnostic to benchmark the performance of computational tools. 

To stabilize the benchmark problems we make several changes to the initial configuration and the model. In particular we replace the well-stirred
 initial dispersion, typically modeled with random initial data, with a fixed bilayer interface configuration with an asymmetric shape and a spatially constant background density of amphiphilic diblock that emulates the reservoir of dispersed molecules. The asymmetry in the shape seeds the motion against curvature. 
 In a benchmark problem this is best not left to random fluctuations as would be the case for a perfectly circular initial shape. 
 For computational reproducibility we smooth the well, replacing the singular well $\Wscmf$ with 
 \begin{equation}
\label{e:Wq-def}
\Wq(u) := \left [ \frac{(u-b_-)^2}{2} +\qtype\,\varepsilon \left (1-\text{sech}\Big (\frac{u-b_-}{\varepsilon}\Big) \right) \right ] \left [ \frac{(u-b_+)^2}{2} +\frac{\gamma}{3} \Big(u- \frac{3b_+-b_-}{2} \Big ) \right ],
\end{equation}
 where the parameter $\qtype$ regulates the second derivative $\Wq''(b_-)$, as depicted in Figure\,\ref{f:W-qtype} (right). This allows a range of approximation of the singularity of the left well of $\Wscmf$.  We fix  $b_\pm=\pm1$ and take the asymmetry parameter $\gamma=0.3$ to match the shape of $\Wscmf$. The perturbative potential $P$ is also singular, and is regularized via replacement with the standard FCH functionalization terms to facilitate comparison to prior analytical results. This yields the non-singular FCH free energy model
\begin{equation}
\label{Functional}
\mE(u):=\int _{\Omega} \frac{1}{2} \Big (\varepsilon^2 \Delta u - \Wq'(u) \Big )^2 - \left (  \frac{\varepsilon^2}{2} \eta_1 |\nabla u|^2+ \eta_2 \Wq(u) \right) \text{d}x,
\end{equation}
where the values of the functionalization parameters $\eta_1$ and $\eta_2$ are determined from a least-square fit of $P$ for the long-chain data. This model fits within the general framework proposed in \cite{GK-93}. All parameter values for each benchmark are recorded in Table\,\ref{t:Benchmark}. For the critical case, the value of $\eta_2$ 
is tuned to enhance the strength of the pearling transient.

The FCH equation is given by the $H^{-1}$ gradient flow of $\mE$
\begin{equation}
\label{sFCH}
u_t = \Delta \frac{\delta \mE}{\delta u},
\end{equation}
which takes the explicit form
\begin{equation}
\label{FCH}
u_t =\Delta \left[  \big (\varepsilon^2 \Delta - \Wq''(u) \big ) (\varepsilon^2 \Delta u - \Wq'(u) ) - \left(  - \varepsilon^2 \eta_1 \Delta u  + \eta_2 \Wq'(u) \right) \right].
\end{equation}

The regularized form of the FCH possesses several advantages. It encompasses both the smooth $\qtype=0$ and the stiff $\qtype>0$ models, naturally allowing for a quantification of the impact of nonlinear stiffness on the computational schemes. While the stiff version mimics the SCMF reduction, the smooth FCH model has been much better studied \cite{CP-19, CP-21, Dai-13, Dai-15} and has advantages in applications which require a simple model that stabilize higher codimensional morphologies with a minimum of numerical stiffness. These applications include the hybrid phase field models for fluid-structure interactions \cite{Qi-Wang2022}.

\subsection{FCH model calibration and benchmark motivation}
\label{s:Cal-BM}

To calibrate the parameters in the regularized well it is convenient to exploit a rescaling of the FCH-SCMF energy that leaves the associated gradient 
flow invariant:
$$ \ep\rightarrow \frac{\ep}{\sqrt{\nu}},\,\, \Wscmf\rightarrow \frac{\Wscmf}{\nu},\,\, P\rightarrow \frac{P}{\nu^2},\,\, t\rightarrow \nu^2 t.$$ 
The rescaling of $\ep$ is equivalent to a change in domain size $L\rightarrow \sqrt{\nu}L.$ 

\begin{figure}[ht!]
\centering
\includegraphics[width=2.3in,trim={5cm 9cm 5cm 8.5cm},clip]{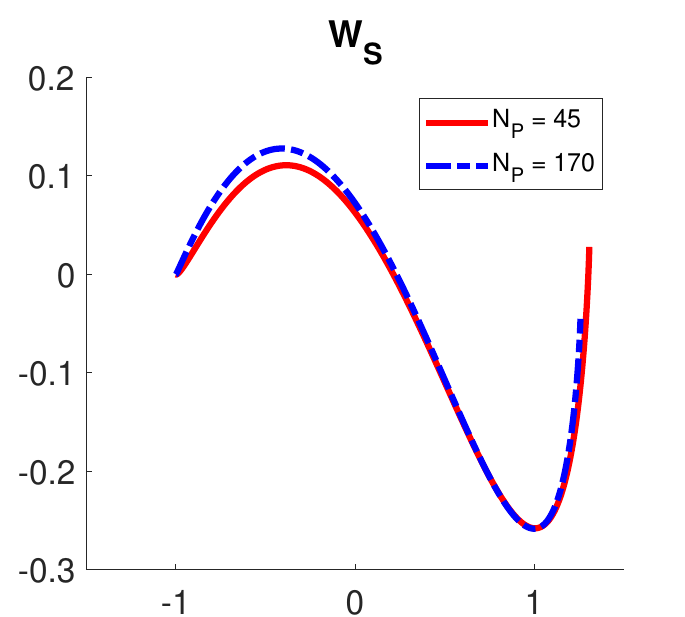} \hspace{0.1in}
\includegraphics[width=2.3in,trim={5cm 9cm 5cm 8.5cm},clip]{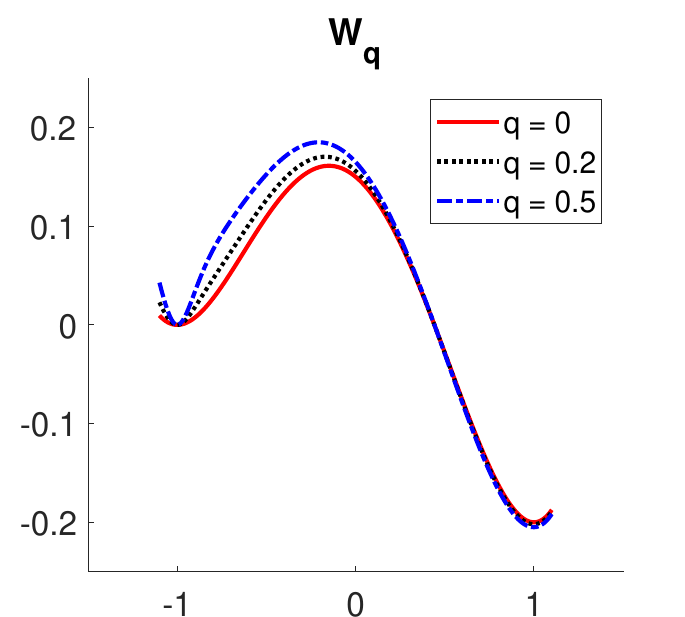} \hspace{0.1in}
\caption{ (left) Graph of scaled singular well $\Wscmf$ as recovered by reduction of SCMF for $N_P=45$ (red) and $N_P=170$ (blue-dotted). (right) Graph  of the regularized well, $\Wq$ for $\qtype=0, ~0.2,~0.5$. }
\label{f:W-qtype}
\end{figure}

We take each monomer to have equal weight, equal to the molecular weight of the solvent. Correspondingly the weight fraction of PEO, $w_{\textrm{PEO}}$, 
equals the molar fraction, $\alpha_A$,  and the polymer weight fraction within the solvent reduces to the molar fraction of polymer, 
$$m_f= \frac{n_PN_P}{n_s} = \frac{1}{100}.$$
For the short-chain polymer benchmark we take $N_P=45$ and $C_0=0.8$ and for the long-polymer benchmark we take $N_P=170$ and $C_0=3.0$,
and rescale the well $\Wscmf$ by a factor of $\nu=4.4$. For the short-chain and long chain polymers the respective choices $b_l=-0.0097$  and $b_l=-0.01+10^{-7}$ sets the left well of $\Wscmf$ at $u=-1$.
The scaled $\Wscmf$ is presented in Figure\,\ref{f:W-qtype} (left) and compared to the regularized well $\Wq$ used in the benchmark simulations.

Intuitively, both a high density of dispersed diblock polymers or a high energy associated to an isolated diblock molecule correspond to a high rate of 
absorption of the dispersed polymers onto the bilayer interface.  The arrival rate is a key quantity controlling defect formation. 
When the arrival rate is slow, the bilayer interface can grow in size to accommodate the new mass. The growth process is adiabatic and has been studied rigorously, \cite{CP-19}, deriving a motion {\sl against} curvature, regularized by a higher order Willmore term that includes surface diffusion. 
If the rate of arrival increases beyond a critical threshold, then defects, such as pearling, endcaps, and loop formation are observed. At moderate rates, a pearling bifurcation can be triggered, the onset of which is well understood within the context of the FCH gradient flow, \cite{NP-19}. The pearling can be transient, subsiding as the dilute suspension of amphiphilic material is consumed. The pearling can also be lead to the formation of end-cap type defects, essentially micelles that remain connected to the underlying structure from which they emerged. The endcaps form most readily at points of high curvature of the bilayer interface. The stem of the endcap can grow, 
forming a long trailing bilayer-type stem and may ultimately reconnect with the initial structure, forming a loop.  At yet higher arrival rates the bilayer interface itself may undergo 
curve splitting -- directly forming closed loops and network structures. The rich array of possible outcomes, and the wide variety of end-states of the gradient
flow, provide an excellent diagnostic of the accuracy of the  proposed schemes.

The benchmark problems introduce two methods to control the rate of arrival of surfactant at the interface. The first is through background level of amphiphilic molecules, controlled by the parameter $\dcoef$ in \eqref{dcoef}, increasing $\dcoef$ corresponds to adding more amphiphilic material to the dispersion. The second is through the convexity of the left well in $W_\qtype$, controlled by the parameter $\qtype$ and the value of $\ep.$ Increasing the value of  $\qtype$ increases $W_\qtype''(-1),$ leading to an increase in the energy of dispersed amphiphilic molecules, which also increases their rate of arrival. The energy of dispersed chains increases with chain length due to the exposure of a longer hydrophobic tail to solvent, \cite{Bates-PRL10}. This is evident within the singular model through the scaling of $\Wscmf$ with $N_P$ via $m_f.$

In the first three benchmarks we take $\qtype=0$, corresponding to shorter chains, and induce bifurcation by raising the background density.
At the low background level in the sub-critical benchmark the initial bilayer interface absorbs amphiphilic material and increases its length, however the rate of absorption is sufficiently slow that there is no generation of defects. In the super-critical benchmark the background level is raised and the elevated rate of arrival induces formation of several defects that coalesce and merge over time.  In the critical benchmark the aspect ratio parameter $\eta_2$ is tuned to extend the duration of the pearling transient within the bilayer interface. Accurate simulations of this benchmark approach the formation an endcap defect before relaxing back to a smooth bilayer profile as the reservoir of dispersed diblock molecules is depleted. In the Foot 1 and Foot 2 benchmarks, we return to the low dispersion level of initial data and systems parameters of the sub-critical case, but increase the value of value of $\qtype$ within the well. This corresponds to lengthening the polymer chains,   increasing the rate of absorption without adjusting the total amount of material absorbed. In both Foot 1 and Foot 2 this induces defect formation.

\subsection{The initial data}
Space is discretized through the standard Fourier pseudo-spectral method assuming periodic boundary conditions on square domains. For the benchmark computations it is useful to have smooth periodic initial conditions on uniform grids. To begin, we fix $\Omega= [0,L]^2$, with $L=4 \pi$, and set the number of grid points along the $x_1$ and $x_2$ axes to be $N_o=256$, corresponding to a mesh spacing $h_o =L/N_o$.  Given a simple  non-intersecting parametric curve $\Gamma=\left\{\big  (x_1(t),x_2(t) \big )\big |~ t_0 \le t \le t_1\right\}$, we construct a region $\Gamma_R$ of uniform width $R$ about $\Gamma$, with outer and inner boundaries $\Gamma_\pm$ defined by
\begin{equation}
\Gamma_\pm=\left\{ \Big (x_1(t) \pm \dfrac{x_2'(t)}{s(t)}R, ~ x_2(t) \mp \dfrac{x_1'(t)}{s(t)}R \Big) \ \middle| \  t_0 \le t \le t_1\right\},
\end{equation}
where $s=s(t)$ is the arc-length of $\Gamma$. We construct the piece-wise constant function $\phi_\Gamma$ to be $1$ inside $\Gamma_R$ and $-1$ outside, and smooth it by convolution
with the filter $\mathbb{F}: L^2(\Omega) \to C_{per}^{\infty} (\Omega)$, defined via
$$\mathbb{F}[\phi_\Gamma] (x) = \sum_{k_1,k_2 \in  I_N} \hat{\phi}_{o,\Gamma} (k_1,k_2)  \exp \left (- \lambda_0 (k_1^2+k_2^2)\right ) \exp \left (\frac{2 \pi i}{L} (x_1 k_1 + x_2 k_2) \right ), $$
where $\hat{\phi}_{o,\Gamma}$ is the discrete Fourier transform (DFT) of $\phi_\Gamma$ interpolated to the $N_o\times N_o$ mesh with spacing $h_o=L/N_o$ and $\lambda_0=7.0269\times10^{-3}$. With the choice $R=0.14725$ the total mass of $\mathbb{F}[\phi_\Gamma]$ per unit length of $\Gamma$ approximates the mass of an exact bilayer dressing of $\Gamma$.  For a fixed curve $\Gamma$ we define $\phi_{256}(x):= \mathbb{F}[ \phi_\Gamma](x)$, which is clearly smooth and $\Omega$-periodic. 

Now, let $N$ be an arbitrary positive integer (typically a power of 2 in the Fourier pseudo-spectral setting), with $h = L/N$. For each of the benchmark cases we define the initial data to be
\begin{equation}
\label{dcoef}
u^0_{N,i,j} = \phi_{256}(ih,jh) +  \varepsilon \frac{\dcoef }{\alpha_m^2(0)}, \quad 0\le i,j\le N,
\end{equation}
where $\dcoef\in\mathbb{R}$ is a parameter that varies in the benchmarks and $\alpha_m(0)=\Wq''(b_-)\bigl|_{\qtype=0}$. Clearly, $u^0_N$ will be a periodic grid function. The curve $\Gamma$ is defined through polar variables as $\Gamma=\Big \{ \big (\rho(\theta)\cos(\theta)+\frac{L}{2}, ~\rho(\theta)\sin(\theta)+\frac{L}{2} \big ) \bigl| \theta \in [0,2\pi) \Big \}$, where
$$\rho(\theta) = 3-\frac{\varepsilon}{2} \cos \big (6(\theta-\frac{\pi}{11}) \big ) -\varepsilon^2 \cos \big (\theta - \frac{3 \pi}{11} \big ).$$

\begin{figure}[ht!]
\centering
\includegraphics[height=6cm,trim={4cm 8cm 4cm 7.5cm},clip]{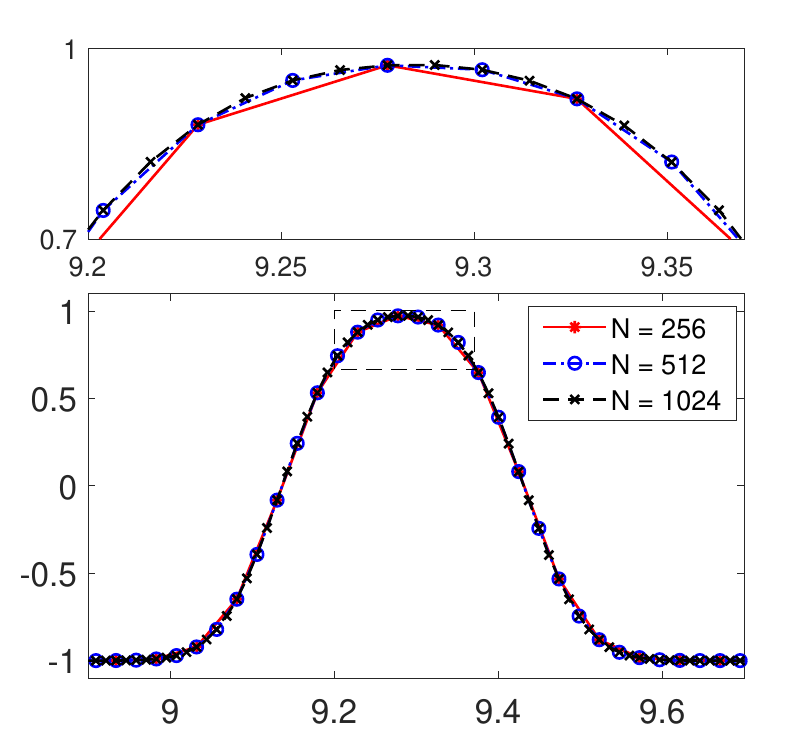}
\includegraphics[height=6cm,trim={0.5cm 4cm 0cm 0.5cm},clip]{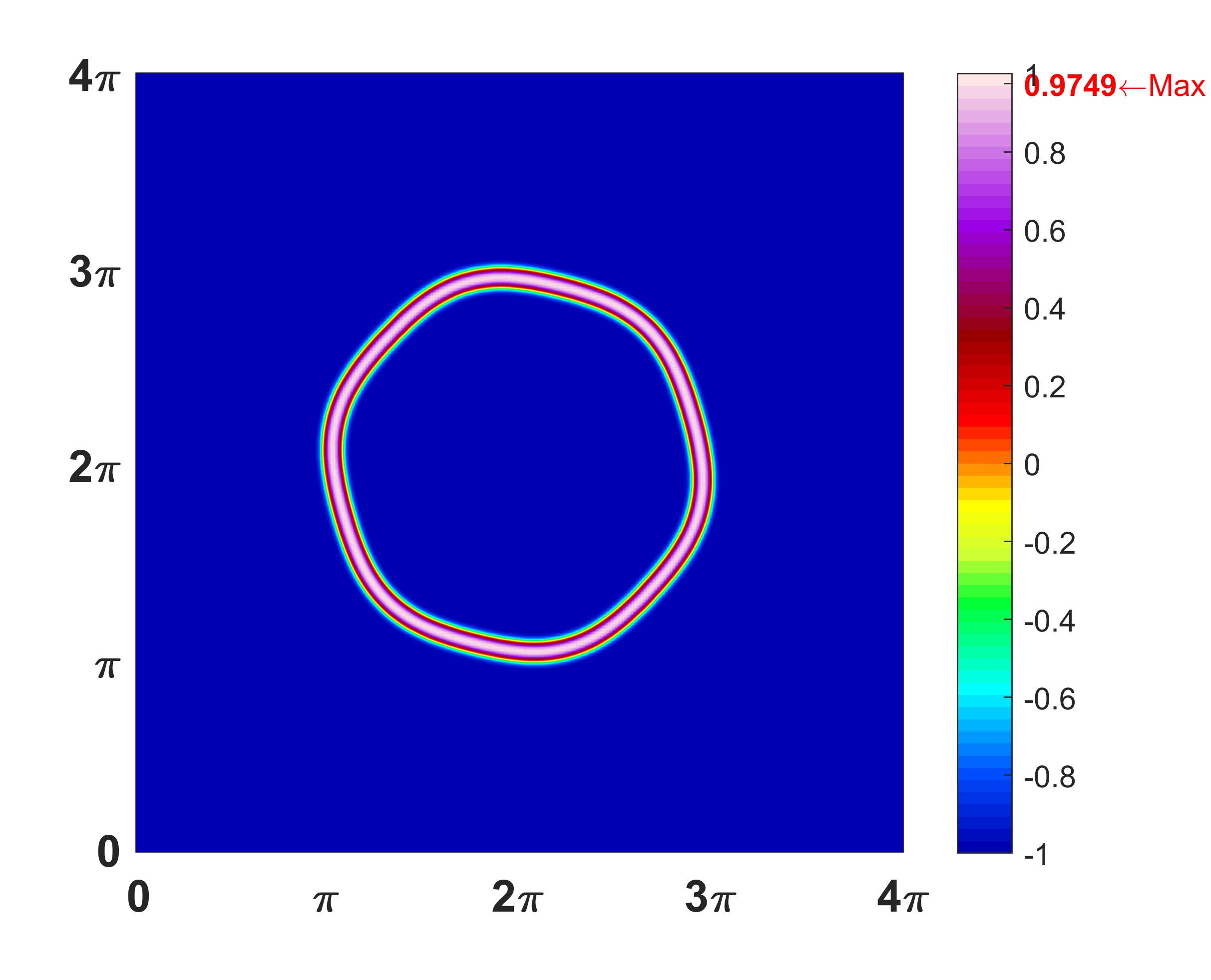}
\label{Init-aSym-SixFold}
\caption{(left) A $1D$ cross-section of the grid function $u^0_{\rm 256}$, along with finer mesh realizations $u^0_{\rm 512}$ and $u^0_{\rm 1024}$. (right) 
    The initial data $u^0_{512}$ constructed from \eqref{dcoef} with width $R=0.14725$ and $\dcoef = 0$. The red number on the colorbar indicates $\max\limits_{i,j} \{ u^0_{512,i,j} \}$.}
\label{Initial}
\end{figure}

The initial data $u^0_N$ corresponding to $N=256$ with this choice of $\Gamma$ is shown in Figure\,\ref{Initial} (right) for $\dcoef=0$. The curve $\Gamma$ is chosen to break any symmetry with the periodic domain and to seed the curvature growth of the bilayer interface.  
The mass, $m_0$, of the initial data, defined via the relation
$$ m_0:= \frac12 \int_\Omega (u^0_N+1)\, \text{d}x,$$
is reported in Table\,\ref{t:Benchmark}.

\begin{table}[ht!]
\centering
\caption{Parameters for Benchmark Cases.}
\renewcommand{\arraystretch}{1.2}
\begin{tabular}{|c|c|c|c|c|c|c|c|c|c|}
\hline
Case$\backslash$Param & $\qtype$ &$\eta_1$  &   $\eta_2$   & $\dcoef$  &  $\ep$ &$\gamma$ & $\alpha_m(\qtype)$ & $N$ &  Mass   \\ \hline 
Sub-critical     &0        &1.45$\ep$& 3$\ep$       & 0.2    & 0.1       & 0.3 &1.7 &256 & 6.11 %0.0387   
\\ \hline
Critical       &0       &1.45$\ep$ & 1.5$\ep$   & 0.75   & 0.1       & 0.3 & 1.7 &256 & 7.61 %0.0482
\\ \hline 
Super-critical  &0       &1.45$\ep$ & 3$\ep$      &  0.5    & 0.1       & 0.3 &  1.7 &256& 6.93 %0.0439  
\\ \hline
Foot 1        &0.2    &1.45$\ep$ & 3$\ep$      & 0.2     & 0.1       & 0.3 & 5.1  &256 & 6.11 %0.0387 
\\ \hline
Foot 2       &0.5    &1.45$\ep$ & 3$\ep$      & 0.2     & 0.1       & 0.3 &  10.2 &512 & 6.11 %0.0387 
\\ \hline
\end{tabular}
\label{t:Benchmark}
\end{table}

\section{The numerical schemes}
As we indicated in the previous section, we use the Fourier pseudo-spectral method to discretize space and simplify the spatial differential operators. The details are standard and skipped for the sake of brevity. In what follows, for simplicity we will write the numerical schemes semi-discretely, using the spatially continuous differential operators, though in practical computations these are replaced by their standard pseudo-spectral approximations.

We use the second order backward differentiation formula (BDF2) to produce the \IMEX\!\!, \PSD\!\!, and \SAV schemes, and use the solution from the third order Adams-Moulton (AM3) scheme as a predictor to control the local error to resolve the benchmark problems described in Section\,\ref{s:Cal-BM}. 

\subsection{Variable step size BDF2 and AM3 schemes}
Consider the initial value problem,  $u'(t)=F(u)$, $u(t^0)=u^0$, for $t_0 \le t \le T$. Let us denote the temporal step size via $k_n:=t^{n}-t^{n-1}$.

Suppose the second order variable step size BDF2 scheme has the form
\begin{equation}
\label{BDF2-general0}
a u^{n+1}+bu^n+c u^{n-1} = F(u^{n+1}),
\end{equation}
where, upon Taylor expanding  and comparing the coefficients, we may identify 
\begin{equation}
\label{BDF2Coef}
\left \{ \begin{aligned}
a &= \frac{1}{k_{n+1}}+\frac{1}{k_{n+1}+k_{n}}     \\
b &= -\frac{1}{k_{n+1}}-\frac{1}{k_{n}}     \\
c &= \frac{1}{k_{n}}-\frac{1}{k_{n+1}+k_{n}}     \\
\end{aligned}. \right.
\end{equation}
Introducing the time-step ratio $\gamma := \frac{k_{n+1}}{k_n}$, the variable step size BDF2 scheme can be written as
\begin{equation}
\label{BDF2-general}
\frac{1+2\gamma}{1+\gamma}u^{n+1}-\frac{(1+\gamma)^2}{1+\gamma} u^n + \frac{\gamma^2}{1+\gamma} u^{n-1} = k_{n+1} F(u^{n+1}),
\end{equation}
which recovers the classical uniform version $3u^{n+1}-4u^n+u^{n-1} = 2k F(u^{n+1})$ when $\gamma = 1$.

Suppose the third order variable step size AM3 scheme has the form
$$u^{n+1} = u^n + \Big [\omega_1 F(u^{n+1}) +\omega_2 F(u^n) +\omega_3 F(u^{n-1}) \Big ].$$
To identify the coefficients $\{\omega_i\}_{i=1}^3$, we make the approximation 
$$u(t^{n+1})-u(t^n) = \int _{t^n} ^ {t^{n+1}} F(u(t)) \text{d}t \approx \int _{t^n} ^ {t^{n+1}} P(t) \text{d}t,$$
where the quadratic polynomial $P(t)$ is the interpolant of $F(u(t))$ at $t^{n-1}, t^n$ and $t^{n+1}$. Therefore the variable step size AM3 is
\begin{equation}
\label{AM3-general}
u^{n+1} = u^n +\frac{k_{n+1}}{6} \left [ \frac{3+2\gamma}{1+\gamma} F(u^{n+1}) + (3+\gamma) F(u^n) -\frac{\gamma^2}{1+\gamma} F(u^{n-1}) \right ],
\end{equation}
which recovers the uniform version $u^{n+1} = u^n +k\Big [ \frac{5}{12} F(u^{n+1}) + \frac{2}{3} F(u^n) -\frac{1}{12} F(u^{n-1}) \Big ]$ when $\gamma = 1$. Further details about these two methods can be found in \cite{Hairer-93}. 
    
\subsection{Adaptive schemes}
The FCH gradient flow \eqref{sFCH}, which may be written as $u_t=F(u)$, where $F(u)=\Delta \frac{\delta \mE}{\delta u}$, undergoes bifurcations that trigger hidden timescales. As these events occur at unpredictable times, an adaptive approach to time-stepping is required to balance accuracy and efficiency.  To initialize the algorithm, we set a target local truncation error tolerance, $\sigma_{\textrm{tol}}$, and the minimal and maximal time-step values $k_{\rm min}$ and $k_{\rm max}.$

Given initial data $u^0$, initial time $t^0$, and some final time $T$, we fix the temporal step size $k_1:=k_{\min}$ and compute the first time-step approximation $u^1$ at time $t^{1}$ for the FCH equation \eqref{sFCH} via an appropriate version of (locally) second order method. 
The adaptive algorithm, based upon \cite{Shen-SIAMRev, Stoer2002}, then proceeds as follows. 
    \begin{description}
    \item[\textbf{Step 0:}]
Given time index $n\in\mathbb{N}_+$, and approximations $u^{n-1}, u^n$ at times $t^{n-1}$ and $t^n$, respectively, with time step sizes $k_n = t^n-t^{n-1}$ and initial $\tilde{k}_{n+1}:=k_n$.
    \item[\textbf{Step 1:}] 
Compute a second order accurate primary approximation $\tilde{u}^{n+1}$ using one of the BDF2 schemes (from the next three sections) using step sizes $k_n$ and $\tilde{k}_{n+1}$.
    \item[\textbf{Step 2:}] 
Compute the time step ratio $\gamma = \frac{\tilde{k}_{n+1}}{k_n}$ and a third order accurate approximation, $u_p$, via the AM3 scheme:
\begin{equation}
\label{Predictor}
u_p := u^n +\frac{\tilde{k}_{n+1}}{6} \Big [ \frac{3+2\gamma}{1+\gamma} F(\tilde{u}^{n+1}) +  (3+\gamma) F(u^{n}) - \frac{\gamma^2}{1+\gamma} F(u^{n-1}) \Big ].
\end{equation}
\item[\textbf{Step 3:}] Calculate the relative error approximation 
$$e_{n+1} := \dfrac{\|\tilde{u}^{n+1}-u_p\|_{L^2}}{\|u_p\|_{L^2}}.$$
%where $\|\cdot \|_F$ is the Frobenius norm.
\item[\textbf{Step 4:}] %\mbox{ }
\textbf{If} $e_{n+1} \le \sigma_{\textrm{tol}}$ or $\tilde{k}_{n+1}=k_{\min}$, \textbf{then}\\
$~\hspace{1cm}~$ Accept the primary approximation, $u^{n+1} = \tilde{u}^{n+1}$.\\
$~\hspace{1cm}~$ Recalculate $k_{n+1} = \max\big\{k_{\min}, \min \{A_{dp}(e_{n+1},\tilde{k}_{n+1}),k_{\max}\}\big \},$
and update the current time, $t^{n+1} = t^n+k_{n+1}$.\\
$~\hspace{1cm}~$ Update the time step index: $n \leftarrow n+1$.\\
$~\hspace{1cm}~$ \textbf{Goto} \textbf{Step 0}.\\
$~\hspace{0.2cm}~$ \textbf{Else} \\
$~\hspace{1cm}~$ Recalculate the time step size $\tilde{k}_{n+1} = \max\big\{k_{\min}, \min\{A_{dp}(e_{n+1},\tilde{k}_{n+1}),k_{\max}\}\big\}$.\\
$~\hspace{1cm}~$ \textbf{Goto} \textbf{Step 1}.\\
$~\hspace{0.2cm}~$ \textbf{Endif}
\end{description}
Here
$$ A_{dp}(e,k) := \rho_s \left(\frac{\sigma_{\rm tol}}{e}\right)^{1/3} k,$$
and we take the safety coefficient $\rho_s = 0.9$, and $k_{\min} = 10^{-9}$ for all simulations.  For the \IMEX and \SAV schemes $k_{\max}$ is taken to be $\infty,$ while for the \PSD scheme, the optimal value of $k_{\max}$ depends upon $\qtype,$ as shown in the Table \ref{tolPSD}. As discussed in \cite{Hairer-93}, to ensure zero-stability for the variable step size BDF2 in \eqref{BDF2-general}, $A_{dp}(e,k)$ needs to be bounded from above by $\big (1+\sqrt{2}\big ) k$. Numerical exploration with this bound on $A_{dp}$ showed it afforded no significant impact on the benchmark problems. 

\begin{remark}
We have chosen the time step adaptivity to directly enforce that the approximate solutions are accurate to a desired local error tolerance, $\sigma_{\rm tol}$. We employ an algorithm similar to that in \cite{Shen-SIAMRev}, though there are several others that have a similar design and purpose, including for example, \cite{Gear1971,Hairer-93,HairerWanner1996,Sonderlind2002}.  The method of \cite{QiaoZhangTang2011} is different in that the energy is monitored in time as a surrogate error indicator. When the preliminary steps indicate a rapid change in energy, the algorithm reduces the time step size with the goal of capturing the corresponding dynamics of the density field, the motivation is that abrupt changes in the energy correspond to topological changes in the density field. A preliminary comparison of the two disparate approaches gives us reason to favor the direct method. First, our objective is accurate density field calculations, and the direct method controls the density field explicitly, rather than implicitly through the energy.  The energy functional is scalar valued, and many classes of deformation do not locally change the value of the energy. This makes the performance of the energy monitoring time-stepping method very sensitive to choices in the time stepping control parameters. Second, the computation of the energy is an added expense that makes the use of an energy-based error indicator less attractive.
\end{remark}

\subsection{The BDF2-PSD scheme}
The BDF2-PSD scheme uses a fully implicit variable time-step BDF2 for the numerical approximation of the system \eqref{FCH} which takes the form
\begin{equation}
\label{BDF2}
au^{n+1}+bu^n+cu^{n-1} = \mG \frac{\delta \mE}{\delta u} \Bigr|^{n+1},
%= \mG \left( \frac{\delta \mathcal{E}}{\delta u} \right)^{n+1}.
\end{equation}
%The BDF2 approximation method may be expressed as
where the coefficients $a,b,c$ are given in \eqref{BDF2Coef}. The solution $u^{n+1}$ in \eqref{BDF2} can be solved in terms of a zero residual,
\begin{equation}
\label{residual}
\mR(u^{n+1}; u^n,u^{n-1}):= \Pi_0\frac{\delta \mE}{\delta u} \Bigl|^{n+1}- \mG^{-1} (a u^{n+1}) - \mG^{-1} (b u^{n}+cu^{n-1}) = 0,
\end{equation}
where $\Pi_0$ denotes the linear zero-mass orthogonal projection operator. 
Given  $u^{n-1}$ and $u^n$, to solve $u^{n+1}$ from \eqref{residual}, this method is accompanied by a preconditioned steepest descent (BDF2-\PSD\!\!) solver, with an approximate line search  (ALS) to invert the highly nonlinear system of equations. This solver is referred to the PSD with ALS, see \cite{CHW-20, FSWW-17}. We refer to this method as \PSD for brevity. 

The preconditioned steepest descent method solves nonlinear system \eqref{residual} iteratively through a series of linear systems. The strictly positive, self-adjoint operator $\mL_{\textrm{PSD}}$ is the linearization of \eqref{residual} about the spatially constant state $u\equiv b_-$ after dropping the small $\eta_1$ and $\eta_2$ terms,
$$\mL_{\textrm{PSD}} := \varepsilon^4 \Delta^2  -2 \alpha_m \varepsilon^2 \Delta  + \alpha_m^2 - a \mG^{-1},$$
which is well-defined on mass-less functions, and preconditions the iterative scheme. Here $\alpha_m=\Wq''(b_-)$ depends strongly on $\qtype$. 
The solution $u^{n+1}$ is thus defined as the limit of the sequence $\{u_s^{n+1}\}_{s=0}^{\infty}$, constructed through the ALS recurrence relation
\begin{align}
\label{recurrence}
u_0^{n+1} &:= u^n+ \tfrac{k_{n+1}}{k_{n}}(u^n-u^{n-1}), \\
\label{recurrence-2}
u_{s+1}^{n+1}&=  u_s^{n+1}+\lambda d^{n+1}_s, \quad s=0,1,2, \ldots
\end{align}
where the search direction $d^{n+1}_s$ at $u_s^{n+1}$ is defined as
$$d^{n+1}_s := - \mL_{\textrm{PSD}}^{-1} \mR(u_s^{n+1},u^n,u^{n-1}).$$
For a prescribed iterative stopping tolerance $i_{\rm tol}$, the ALS procedure is terminated once $\frac{\|d^{n+1}_{s}\|_{L^2}}{\|u^{n+1}_{s+1}\|_{L^2}} < i_{\rm tol}$. The parameter $\lambda$ in \eqref{recurrence-2} is the \text{search-step-size}. 
Numerical investigations show that the optimal value of $\lambda$ is somewhat sensitive to the value of $\alpha_m=\alpha_m(\qtype)$ and temporal step size $k$. This dependence is determined by minimizing the average number of \PSD iterations for a fixed $k$ over the first 50 temporal steps of the simulation.  Optimal values of $\lambda$ for different values of $\qtype$ and $k$ are reported in Table\,\ref{SearchSize}. The values used in the simulations are determined by linear interpolation. 

\begin{table}[ht!]
\centering
\caption{Dependence of optimal value of search-step-size $\lambda$ on temporal step size $k$.}
\renewcommand{\arraystretch}{1.2}
\begin{tabular}{|c|c|c|c|c|c|c|c|c|c|c|c|}
\hline
\diagbox[height=2.8em]{\hspace{-0.15cm}$\qtype$}{\hspace{-0.1cm}$\lambda$}{\hspace{0.1cm}$k$} & $\le 10^{-6}$   & $10^{-5}$ &     $5\cdot 10^{-5}$ &  $10^{-4}$ &  $5\cdot  10^{-4}$ & $10^{-3}$ & 0.005 & 0.01  & 0.02 & 0.03\\ \hline % \hline 
$0$ & 1 & 1.07 & 1.11 & 1.14 & 1.24 & 1.34 & 1.60 & 1.738 & 1.804 & 1.855\\ \hline
$0.2$  & 1 & 1.04 & 1.15 & 1.28 & 1.50& 1.70 & 1.87 & 1.92 & 1.95 & 1.97\\ \hline 
$0.5$  & 1 & 1.20 & 1.32 & 1.45 & 1.72 & 1.83 & 1.965 & 1.97 & 1.985 & 1.99\\ \hline
\end{tabular}
\vskip 0.05in
\label{SearchSize}
\end{table}
% & $5\cdot 10^{-6}$ & 1.05 & 1.01 & 1.16

The iterative stopping tolerance, $i_{\rm tol}$, impacts the accuracy and computational cost of the \PSD scheme.  Numerical optimization finds that an optimal choice of  $i_{\rm tol}$ is sensitive to both the well stiffness, $\qtype$, and the local truncation error, $\sigma_{\rm tol}$. We determine this relation through the ratio
$$  i_{\rm tol}=\nu(\qtype) \sigma_{\rm tol},$$
and determine an optimal value of $\nu(\qtype).$ This requires balance, as overly small values of $i_{\rm tol}$ lead to excessive iterations that do not improve the scheme's accuracy.
On the other hand $i_{\rm tol}$ must be small enough to ensure that numerical error from the iterative solver does not pollute the adaptive time-stepping and does not impede the convergence of the iterative solver at subsequent time-steps. Instructively, the iterative convergence rate is found to depend upon the upper limit, $k_{\max}$, imposed on the adaptive time-stepping algorithm.   
This leads to a coupled numerical optimization study, presented in Table\,\ref{tolPSD}  which shows the sensitively of iterations numbers upon $k_{\max}$ for the three values of $\qtype$, and the optimal value of $\nu$. The iteration counts increase considerably with $\qtype$, while $\nu$ decreases exponentially with $\qtype$. If the upper bound $k_{\max}$ is removed then the iteration count may increase considerably, with associated increase in computational effort. The tuning of $k_{\rm max}$ and $\nu$ with $\qtype$ is the most unpredictable element of the optimization process for any of the schemes.

% Table generated by Excel2LaTeX from sheet 'Convergence_Rate'
\begin{table}[ht!]
\centering
\caption{Dependence of \PSD iteration count on $\qtype$, $\nu$ and $k_{\max}$.}
\renewcommand{\arraystretch}{1.1}   % set the height of rows
\begin{tabular}{|c|c||c|c|c|c|c|c|c||c|}
\hline
\multicolumn{1}{|m{1.8cm}|}{\multirow{2}[4]{*}{\makecell[c]{Iteration\\ count/1000}}} & \multirow{2}[4]{*}{$\nu$(q)} & \multicolumn{7}{c||}{Value of $k_{\max}$}                & \multicolumn{1}{c|}{\multirow{2}[4]{*}{\makecell[c]{optimal\\ $k_{\max}$}}} 
\bigstrut\\
\cline{3-9}          &       & 0.009 & 0.01  & 0.02  & 0.03  & 0.04  & 0.05  & 0.06  &  \bigstrut\\
\hline
q = 0 & 1.E-03 &       & 36.3  & 34.8  & 36.5  & 36.7  & {\bf 38.2}  & 41.4  & 0.05 \bigstrut\\
\hline
q = 0.2 & 2.E-05 &       & 43.3  & {\bf 42.8}  & 43.7  &       &       &       & 0.02 \bigstrut\\
\hline
q = 0.5 & 1.E-06 & 162.0  & {\bf 161.7}  & 162.0  &       &       &       &       & 0.01 \bigstrut\\
\hline
\end{tabular}%
\vskip 0.05in
  \label{tolPSD}
\end{table}%

\subsection{The BDF2-IMEX scheme}
The FCH equation \eqref{FCH} can be rewritten as
\begin{equation}
\label{IMEX-0}
\begin{aligned}
u_t &= \mG \Big[ \mL_{\textrm{IMEX}} u + \cN_{\textrm{IMEX}}(u) \Big],\\
%Sulin: + \varepsilon^2 \eta_1 \Delta u - \eta_2 \Wq'(u) 
%\varepsilon^4  \Delta^2 u - \varepsilon^2 \alpha_m \Delta u + \varepsilon^2 \eta_1 \Delta u + 
%\varepsilon^2 \big  (\alpha_m-W''(u) \big ) \Delta u - \varepsilon^2 \Delta W'(u)+ W''(u)W'(u) - \eta_2 W'(u)\Big],\\
\end{aligned}
\end{equation}
where we introduce the linear positive operator
\begin{equation}
\label{e:IMEX-op}
\mL_{\textrm{IMEX}} :=\varepsilon^4  \Delta^2 -2\alpha_m \varepsilon^2 \Delta + \alpha_m^2,
\end{equation}
obtained by linearizing $\frac{\delta \mE}{\delta u}$ in \eqref{FCH} about $u=b_-$ and dropping the small, negative $\eta_1$ and $\eta_2$ terms. 
The term $\cN_{\textrm{IMEX}}$ is genuinely nonlinear with zero linearization about $u=b_-$, including the $\eta_1$ and $\eta_2$ terms
%Sulin
$$\begin{aligned}
\cN_{\textrm{IMEX}}(u) ~:=~ & \varepsilon^2 \big  (\alpha_m -\Wq''(u) \big ) \Delta u + \varepsilon^2 \Delta \big (\alpha_m u- \Wq'(u) \big ) + \Wq''(u)\Wq'(u) \\
&-\alpha_m^2u + \varepsilon^2 \eta_1 \Delta u - \eta_2 \Wq'(u) .
\end{aligned}
$$
The resulting second order semi-implicit \IMEX scheme is chosen to stabilize the spatially constant background state $u\equiv b_-$. To this end we take the dominant linear terms implicit and the remainder explicit, 
\begin{equation}
\label{IMEX}
\begin{aligned}
au^{n+1}+bu^n+cu^{n-1}
& = \mG \Big[ \mL_{\textrm{IMEX}} u^{n+1}+ \cN_{\textrm{IMEX}}(u^{*,n+1})\Big],
%Sulin:  + \varepsilon^2 \eta_1 \Delta u^{*,n+1} -\eta_2 \Wq'(u^{*,n+1})
%& \hspace{1cm} + \varepsilon^2 \big  (2\alpha_m-W''(u^{n}) +\eta_1 \big ) \Delta u^n - \varepsilon^2 \Delta W'(u^n) + W''(u^n)W'(u^n)-\alpha_m^2u^n  - \eta_2 W'(u^n)\Big].
\end{aligned}
\end{equation}
where $u^{*,n+1}$ can be chosen as any explicit (locally) second order approximation of $u(t^{n+1})$ to make the scheme consistent, for instance,
\begin{equation}
\label{Richardson}
u^{*,n+1} = u^n+ \tfrac{k_{n+1}}{k_{n}}(u^n-u^{n-1}).
\end{equation}
Now we can isolate and solve $u^{n+1}$ in \eqref{IMEX} from
\begin{equation}
\label{IMEX-eq}
\big (a- \mG \mL_{\textrm{IMEX}}\big ) u^{n+1} = -bu^n-cu^{n-1} + \Delta\cN_{\textrm{IMEX}}(u^{*,n+1}).
\end{equation}
%Sulin: + \varepsilon^2 \eta_1 \Delta u^{*,n+1} -\eta_2 \Wq'(u^{*,n+1})

\subsection{The BDF2-SAV scheme}
Computational schemes based upon the \SAV formulation have been applied to the FCH gradient flow, see \cite{ZOWW}. The version presented here is a slight variation. We rewrite the FCH energy functional $\mE(u)$ in \eqref{Functional} in the form:
\begin{align}   
\label{4.1}
\mE(u) = \int_\Omega \left[ \frac{\ep^4}{2} (\Delta u)^2 - \left(\frac{\eta_1}{2}+\zeta\right)\varepsilon^2 |\nabla u|^2 + G(u) \right] \text{d}x,
\end{align}
where $\zeta>0$ is a parameter and
\begin{equation}
\label{e:G-def}
 G(u):= - \varepsilon^2\Delta u \left(\Wq'(u)+\zeta u\right) + \frac{1}{2} (\Wq'(u))^2  -\eta_2 \Wq(u) .
\end{equation}
The choice of principle linear operator for the \SAV scheme is a bit less intuitive than for the \IMEX or \PSD schemes. We introduce
\begin{equation}
\label{e:def-LSAV}
\mL_{\textrm{\SAV}} = \varepsilon^4 \Delta^2 + \varepsilon^2\left(\eta_1+2\zeta\right)\Delta =  \mL_0+ \mL_1,
\end{equation}
where the sub-operators are parameter dependent
\begin{align}
\label{e:SAV-L0}
\mathcal{L}_0(\beta_1,\beta_2)& = \varepsilon^4 \Delta^2 - \beta_1\alpha_m\varepsilon^2\Delta + \beta_2\alpha_m^2,
\\
\label{e:SAV-L1}
\mathcal{L}_1(\beta_1,\beta_2) &= \varepsilon^2\left(\eta_1 +2\zeta \right)\Delta +\beta_1\alpha_m \varepsilon^2 \Delta
-\beta_2\alpha_m^2,
\end{align}
where $\alpha_m = \alpha_m(\qtype)$ and the constants $\beta_1, \beta_2\geq 0$ are the stabilization parameters. 
The operator $\mL_0$ defines the principle linear implicit terms in the \SAV scheme. The default choice for these parameters is 
$\beta_1=2$ and $\beta_2=1.$
% $\beta_1=0$ and $\beta_2=3.$  % this is the original default

Introducing the auxiliary energy
\[ \cE_1(u) = \int_\Omega G(u) \text{d}x,\]
the FCH energy \eqref{4.1} takes the form 
\begin{align}   
\label{4.2} 
\mE(u) = \frac{1}{2} \left(u, \mL_{\textrm{SAV}}\,u\right)_{L^2(\Omega)} + \mathcal{E}_1(u).
\end{align} 

For fixed time-steps the \SAV scheme is known to be energy stable for a modified energy, if the functional $\cE_1(u)$ can be shown to be uniformly bounded from below over $H^2_{\rm per}(\Omega)$, \cite{SXU-18}. This is achieved by choice of $\zeta=\zeta(\qtype).$ Specifically 
$$
\mathcal{E}_1(u) 
\geq \int_{\Omega} \Big ( \Wq''(u)+\zeta \Big ) |\nabla u|^2 \text{d}x + \int_{\Omega} \Big [\frac12(\Wq'(u))^2- \eta_2 \Wq(u)\Big ] \text{d}x,
$$
and choosing $\zeta$ larger than the negative of the minimum value of the $\Wq''$, we estimate
$$
\cE_1(u) \geq  |\Omega| \min_u \Big (\frac12 (\Wq'(u))^2-\eta_2 \Wq(u) \Big ) > -D_0, 
$$ 
where $D_0>0$ only depends upon the domain $\Omega$, the value of $\eta_2$ and $\Wq$.

For the energy splitting approach, we introduce the scalar auxiliary variable 
$$r=r(t) := \sqrt{\mathcal{E}_1(u)+D_0},$$ 
then the FCH equation can be rewritten as
\begin{align}  
\label{4.3}
& \frac{\partial u}{\partial t}= \mG \mu,~~~~ \mu:=\mL_{\textrm{SAV}}\,u + \frac{r~V[u]}{\sqrt{\mathcal{E}_1(u)+D_0}},  \\[1 \jot]
\label{4.4}
&  \frac{d r}{d t}=\frac{1}{2\sqrt{\cE_1(u)+D_0}} \int_{\Omega} V[u] \frac{\partial u}{\partial t} \text{d}x,
\end{align}
where $V[u] = \delta \cE_1/\delta u=G'(u)$. For choosing $u^{*,n+1}$ as in \eqref{Richardson}, the \SAV scheme takes the form
\begin{align}  
\label{4.7}
& au^{n+1}\!+b u^n +c u^{n-1} \!= \mG  \mu^{n+1},~~ \mu^{n+1} \!=\! \mathcal{L}_0 u^{n+1} \!+\! \mathcal{L}_1 u^{*,n+1} \!+\! \frac{r^{n+1}~V[u^{*,n+1}]}{\sqrt{\mathcal{E}_1(u^{*,n+1})+D_0}}, \\[1 \jot]
\label{4.8}
&  ar^{n+1}+b r^n + c r^{n-1} =  \int_{\Omega} \frac{V[u^{*,n+1}]}{2\sqrt{\mathcal{E}_1(u^{*,n+1})+D_0}} \big  (au^{n+1}+bu^n+cu^{n-1} \big ) \text{d}x.
\end{align}

We remark that the $r^{n+1}$ variable in \eqref{4.8} also contributes to the implicit equation for $u^{n+1}$. The full resolution of $u^{n+1}$ from \eqref{4.7}-\eqref{4.8} is presented in \cite{Shen-SIAMRev, ZOWW}, but is driven by  the inversion of the operator $\mL_0-a\Delta^{-1}$. With a fixed time-step $k$, the \SAV scheme is unconditionally energy stable for the auxiliary energy
$$\begin{aligned}
\cE_{\rm aux}\big(u^{n},u^{n-1},r^{n},r^{n-1}\big)
&:= \frac{1}{2}\left (u^n,\mathcal{L}_{\textrm{SAV}} u^n\right)_{L^2(\Omega)} - \left(u^{n}-u^{n-1}, \mathcal{L}_1(u^{n}-u^{n-1})\right)_{L^2(\Omega)}\\ 
&~~~~~ +\frac{1}{2}\left(2u^{n}-u^{n-1} , \mathcal{L}_{\textrm{SAV}}(2u^{n}-u^{n-1})\right)_{L^2(\Omega)} + \big |r^n\big|^2 + \big|2r^{n}-r^{n-1}\big|^2.
\end{aligned}$$
\begin{theorem}
\label{t:SAV}
When implemented with a fixed time step size $k>0$, the \SAV scheme \eqref{4.7}-\eqref{4.8} is unconditionally modified-energy stable in the sense that the discrete modified-energy law holds,
\begin{align}
\label{EnergyProperty-SAV}
\cE_{\rm aux}\big(u^{n+1},u^{n},r^{n+1},r^{n}\big) \leq 
\cE_{\rm aux}\big(u^{n},u^{n-1},r^{n},r^{n-1}\big),~ n\geq 1,
\end{align}
\end{theorem}
The proof of Theorem\,\ref{t:SAV} is given in Appendix A. Details on energy stability properties of \SAV schemes can be found in \cite{Shen-SIAMRev, ZOWW}.

\begin{figure}[ht!]
\centering
\includegraphics[width=4.6cm,trim={3.5cm 8cm 4.5cm 8cm},clip]{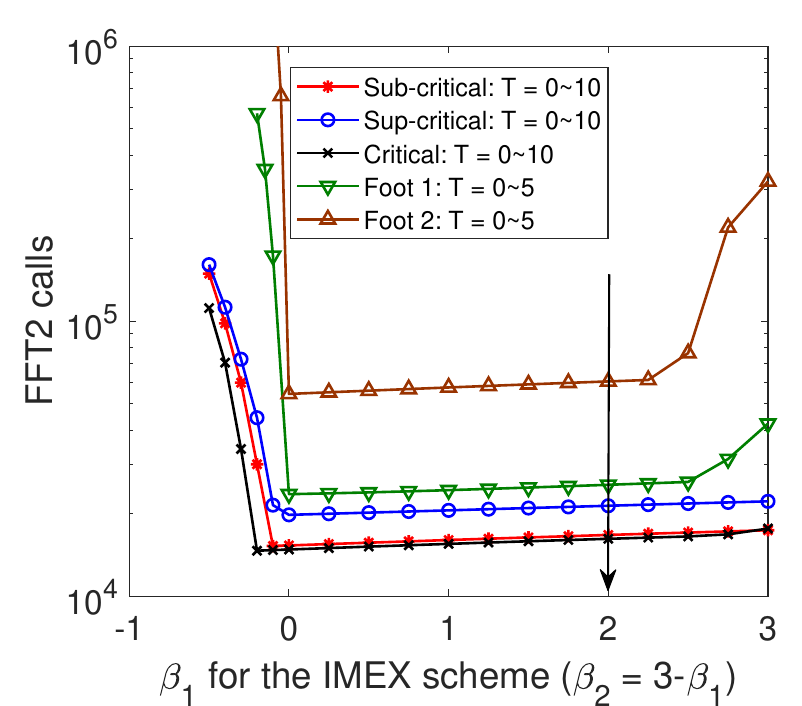}
\includegraphics[width=4.6cm,trim={3.5cm 8cm 4.5cm 8cm},clip]{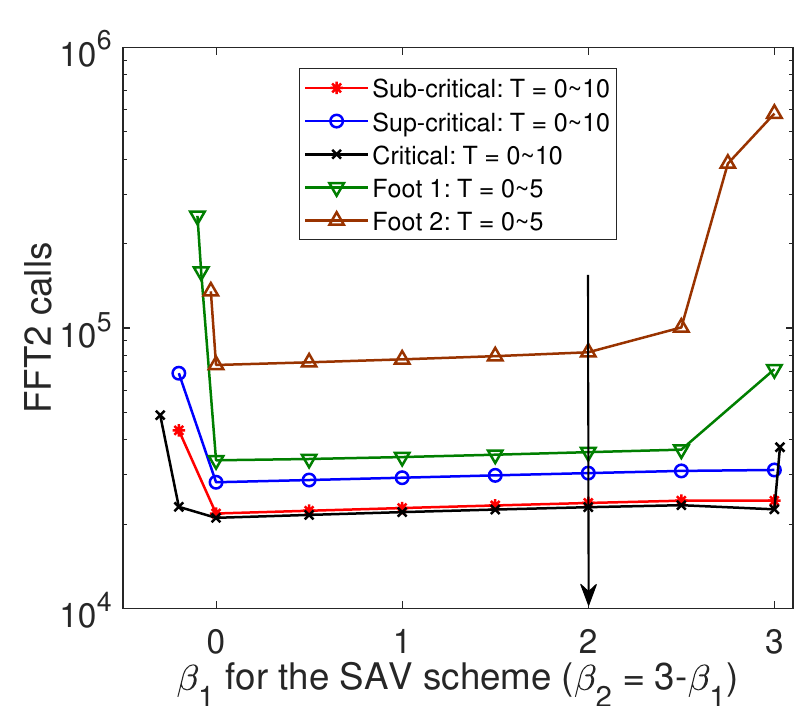}~
\includegraphics[width=4.6cm,trim={3.5cm 8cm 4.5cm 8cm},clip]{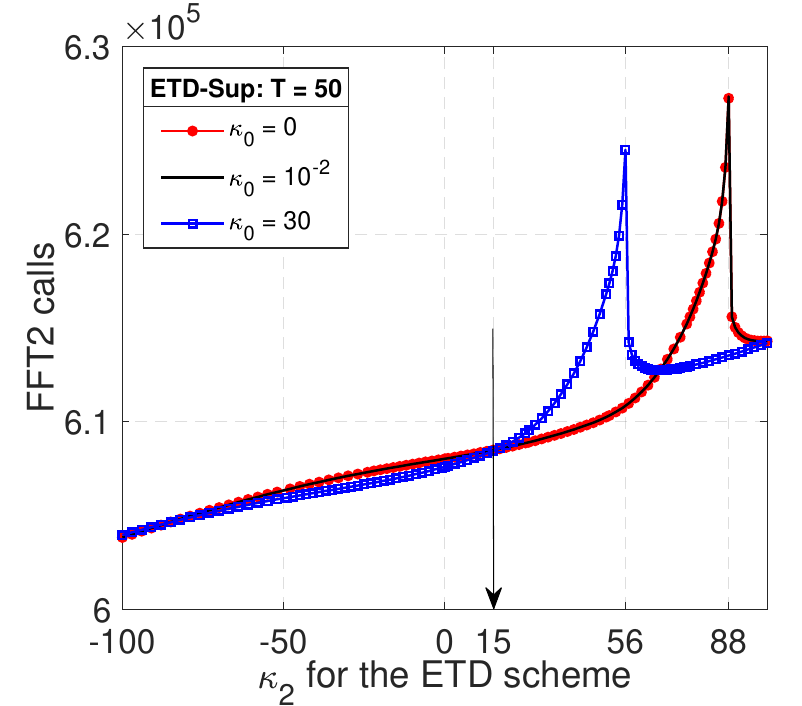}
\caption{ (left/center) Total FFT calls on log scale at time $T=10$ verses stabilization parameters $\beta_1$ with $\beta_2=3-\beta_1$ for \IMEX and \SAV for each of the 5 benchmark simulations.  (right) Total FFT calls on linear scale at time $T=50$ verses stabilization parameter $\kappa_2$ for \ETD for supercritical benchmark and three choices of $\kappa_0.$ The peaks in FFT calls correspond to onset of a shape bifurcation that generates an extra endcap in the \ETD simulation. The black arrow indicates choice of parameters in simulations of Section 4.}
\label{Find-Beta}
\end{figure}

The stabilization parameters make $\mL_0$ a strictly positive operator and play an essential role in the convergence, accuracy, and efficiency of the \SAV scheme. 
The operator $\mL_0$ agrees with $\mL_{\textrm{IMEX}}$ for the choice $\beta_1=2$ and $\beta_2=1$ that we take here. Figure\,\ref{Find-Beta} shows FFT counts for simulations of \IMEX and \SAV using the dominant implicit term based on $\mL_0.$  
Overall the schemes preform well if $\beta_1+\beta_2=3$, with performance deteriorating dramatically for smaller values and slowly for larger values of this sum. Indeed values of $\beta_1+\beta_2<3$ can lead to FFT counts that are several orders of magnitude higher per time-unit at a fixed local truncation error. The left panel provides total FFT counts for the 
\IMEX scheme with $\beta_2=3-\beta_1$, showing that the performance is optimal so long as neither $\beta_1$ nor $\beta_2$ are too small. The right panel shows performance of the \SAV scheme  for each of the five benchmark problems.
The choice $\beta_1=2$ is taken as the default for both \IMEX and \SAV\!\!.

\subsection{The ETDRK2 scheme}
For the temporal discretization, the FCH equation \eqref{FCH} can be viewed as an infinite dimensional ODE written in the following operator splitting form
\begin{equation}
\label{ODE}
\frac{\md u}{\md t} = \mL u+ \N(u),
\end{equation}
where $\mL$ is a negative-definite linear differential operator and $\N (u)$ is remaining nonlinearity. We may multiply both sides of \eqref{ODE} by the linear semigroup $\me^{-\mL t}$, to obtain the  ``exponentiated'' form of \eqref{ODE}
\begin{equation}
\label{ODEeqvi}
\big (\me^{-\mL t} u \big )_t = \me^{-\mL t}\N(u).
\end{equation}

For time-step $k = t^{n+1}-t^{n}$, integrating \eqref{ODEeqvi} over $[t^{n},t^{n+1}]$ yields the time-differenced system,
\begin{align}
\label{ETD}
u(t^{n+1}) &= \me^{\mL (t^{n+1}-t^{n})}u(t^{n}) + \int_{t^n}^{t^{n+1}} \me^{\mL (t^{n+1}-s)}\N(u(s)) \md s,\\
&= \me^{\mL k}u(t^{n}) + \int_{0}^{k} \me^{\mL (k-s)}\N(u(t^n + s)) \md s,
\end{align}
The exponential time differencing (ETD) approach uses this formulation, arriving at an iterative scheme by approximating the integrals with finite differences, 
%$\N \big (u(t^n + s) \big )$ for $s \in [0,k]$ and evaluate the integrals about $s$ exactly, see 
more details can be found in \cite{Cox-02,DJLQ-18,DJLQ-21,JZZD-15}. Precisely, the explicit first order ETD Runge-Kutta (ETDRK1) scheme uses the approximation $\N \big (u(t^n + s) \big ) \approx \N(u^n)$ for $s \in [0,k]$. This yields
\begin{equation}
\label{ETDRK1}
u^{n+1} = \me^{\mL k}u^{n} + \int_{0}^{k} \me^{\mL (k-s)} \md s ~ \N(u^n) = \me^{\mL k}u^{n} + \mL^{-1} \big (\me^{\mL k}-I\big ) \N(u^n).
\end{equation}

The explicit second order ETD Runge-Kutta (ETDRK2) scheme uses %a trapezoidal approximation for the integral, equivalently 
a linear approximation for 
$\N \big (u(t^n + s) \big ) \approx (1-\frac{s}{k})\N(u^n)+\frac{s}{k}\N(u^{n+1})$ for $s \in [0,k]$. This yields the scheme
\begin{equation}
\left \{ \begin{aligned}
{\color{blue}\tilde{u}^{n+1}} &= \me^{\mL k}u^{n} + \int_{0}^{k} \me^{\mL (k-s)} \N(u^n) \md s,\\
u^{n+1} &= \me^{\mL k}u^{n} + \int_{0}^{k} \me^{\mL (k-s)} \left [  (1-\frac{s}{k})\N(u^n)+\frac{s}{k}\N({\color{blue}\tilde{u}^{n+1}}) \right ] \md s.
\end{aligned} \right. 
\end{equation}
Evaluating the integrals exactly, we find
\begin{equation}
\label{ETDRK2}
\left \{ \begin{aligned}
{\color{blue}\tilde{u}^{n+1}} &= \me^{\mL k}u^{n} + \mL^{-1} \big (\me^{\mL k}-I\big ) \N(u^n),\\
u^{n+1} 
%&= \me^{\mL k}u^{n} + \mL^{-1} \big (\me^{\mL k}-I\big ) \N(u^n) + \left [ \mL^{-2} (\me^{\mL k}-I) -k \mL^{-1} \right ] \frac{\N({\color{blue}\tilde{u}^{n+1}})-\N(u^{n})}{k} \\
&= {\color{blue}\tilde{u}^{n+1}} + \mL^{-1} \left [ \mL^{-1} (\me^{\mL k}-I) -k I \right ] \frac{\N({\color{blue}\tilde{u}^{n+1}})-\N(u^{n})}{k}. \\
\end{aligned} \right. 
\end{equation}

For the FCH equations \eqref{FCH} and \eqref{ODE}, we mirror the \IMEX and \SAV approach, choosing
\begin{equation}
\label{ETD-L}
\left \{ \begin{aligned}
\mL &= \Delta \Big (\varepsilon^4  \Delta^2 -2\alpha_m \varepsilon^2 \Delta + \alpha_m^2 \Big ) -\kappa_0 \mathcal{I} + \kappa_2 \varepsilon^2 \Delta, \\
\N(u) &= \Delta \N_{IMEX}(u) + \kappa_0 u - \kappa_2 \varepsilon^2 \Delta u ,
\end{aligned} \right. 
\end{equation}
%Sulin: we use the general form of ETDRK2 above and below
where $\kappa_0,\kappa_2$ are some positive constants and $\mathcal{I}$ is the identity operator. Numerical tests show that FFT calls of ETDRK2 are not sensitive to the choice of $\kappa_0$ and $\kappa_2 $. 
%However, accuracy improves slightly for choices of $\kappa \in [10^{-4}, 10]$. 
We take $\kappa_0 = 10^{-2},~\kappa_2 = 15$ in all simulations because of slightly improvement for FFT calls and accuracy. We refer to ETDRK2 as \ETD for brevity.

\section{Benchmark simulations}
\label{Simulations}
We present an overview of the Benchmark simulations for local truncation error $\sigma_{\rm tol}=10^{-5}$, for which the \PSD scheme is accurate while the \IMEX\!, \SAV\!, and \ETD schemes are borderline accurate.  Generically we find that a global $L^2(\Omega)$ relative discretization error of $2.5\times10^{-3}$ is  sufficient to ensure that each scheme is quantitatively accurate, with the correct numbers, types, and placements of defects.

\begin{table}[htbp]
  \centering
  \caption{$L^2$ relative error between (\PSD\!, \IMEX\!, \SAV\!, ETD) for each benchmark simulations at final time.} 
  \renewcommand{\arraystretch}{1.2}
    \begin{tabular}{|c|c|c|c|c|c|}
   % \multicolumn{3}{c}{{L2 Error - Compared with larger N}} \\
\hline
   Benchmark & \IMEX$\!\!$/\PSD & \SAV$\!\!$/\PSD & \SAV$\!\!$/\IMEX & ETD/\PSD & $T$\\ \hline
    {Sub-Critical} & 7.276E-03 &  7.315E-03 &  4.103E-05 & & 250  \\ \hline
    {Critical} & 2.204E-03 &  2.212E-03 &  1.796E-05 & & 250 \\ \hline
    {Super-Critical} & 8.817E-02 &  8.819E-02  & 4.156E-05 &  1.568E-02 & 250 \\ \hline 
    {Foot 1} & 2.358E-03 &  2.359E-03  &  1.702E-06 & & 50 \\ \hline
    {Foot 2} & 3.318E-04 &  3.322E-04  &  6.195E-07 & & 50 \\ \hline
    \end{tabular}%
  \label{t:Comp-error}%
\end{table}%

\subsection{Sub-critical benchmark}
\begin{figure}[htbp!]
\centering
\includegraphics[width=13.6cm,trim={0.75cm 0.15cm -0.15cm 0.5cm},clip]{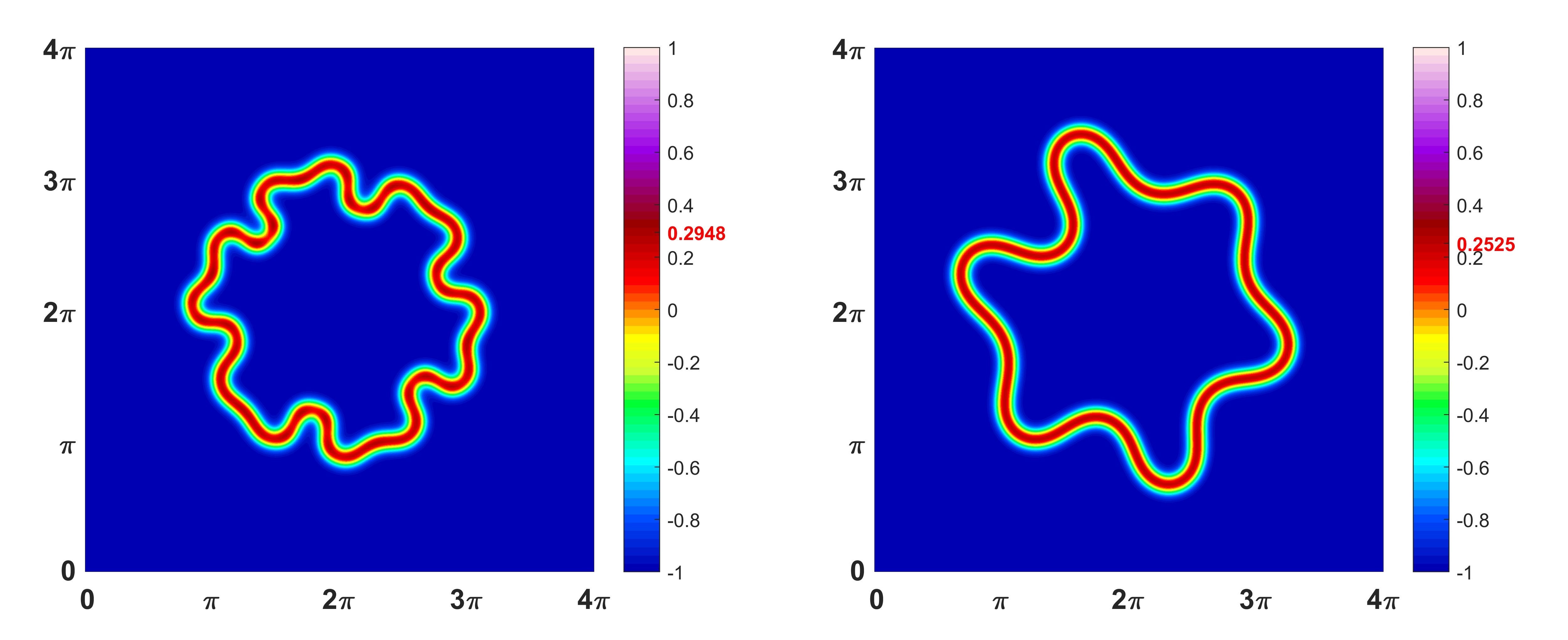}
\vspace{-0.1in}
\caption{Simulation of the sub-critical benchmark with $\qtype=0$, $\sigma_{\rm tol}=10^{-5}$ and $N=256$ at times $T=10$(left) and $T=250$(right).
 All schemes agree to within $L^2$ relative error $7\times10^{-3}$ as reported in Table\,\ref{t:Comp-error}. The red number on colorbar indicates $\max \{u\}$.}
\label{f:BM1}
\end{figure}

The sub-critical benchmark has a low level of dispersed diblock 
polymer material, controlled by the parameter $\dcoef$ in \eqref{dcoef}, while the relatively mild concavity of $\Wq$ at $u=b_-$, controlled by $\alpha_m(0)=\Wq''(b_-)\bigl|_{\qtype=0}$, leads to a gentle absorption rate. The bilayer interface profile does not pearl and remains a simple closed curve from initial data to its final equilibrium shape, as shown in Figure\,\ref{f:BM1} at times $T=10$ and $T=250$.
As shown in \cite{CP-19}, gentle absorption drives motion against curvature, regularized by surface diffusion, which relaxes to a curvature driven flow as the background material is depleted.  All schemes are in quantitative agreement,  as can be verified by the contour plot comparison in Figure\,\ref{f:contour} (left) and the data of Table\,\ref{t:Comp-error}.

\subsection{Critical benchmark} 
\begin{figure}[ht!]
\vspace{-0.1in}
\centering
\includegraphics[width=13.6cm,trim={0.75cm 0.15cm -0.15cm 0.5cm},clip]{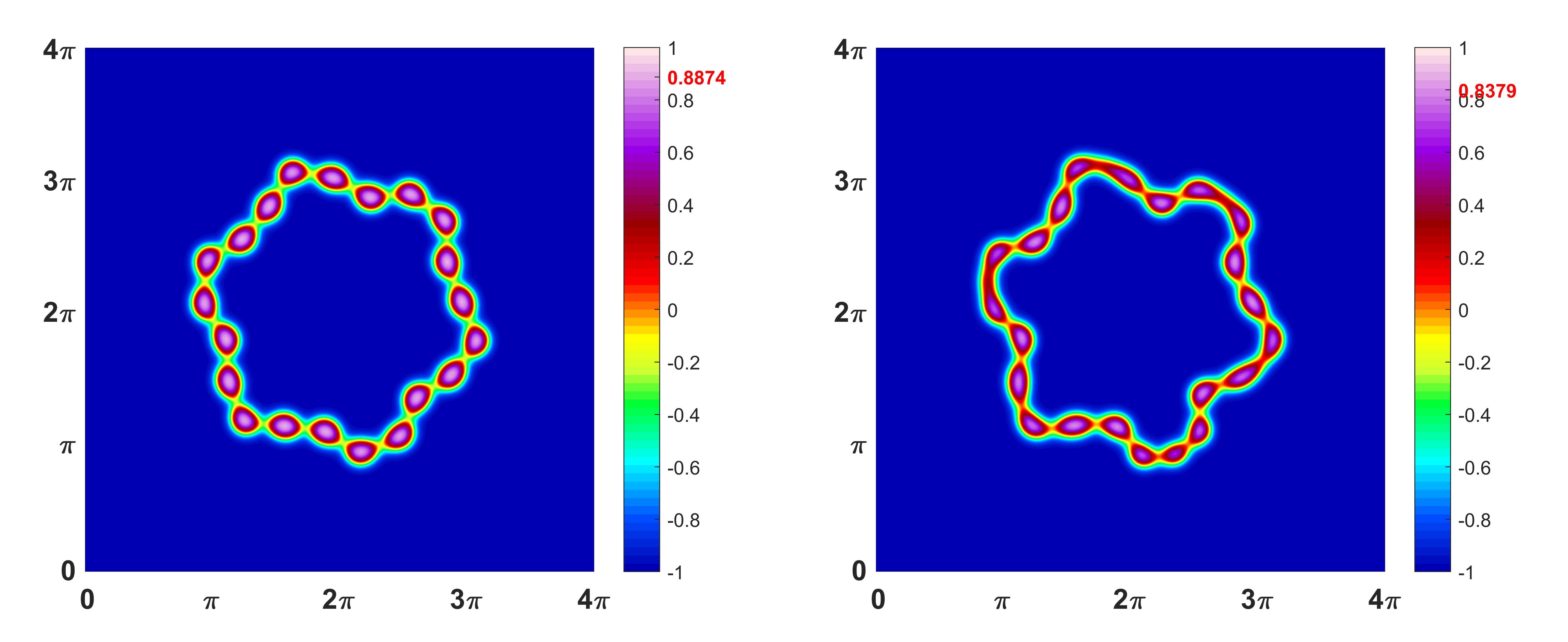}
\vspace{-0.1in}
\caption{Simulation of the critical benchmark with $\qtype=0$, $\sigma_{\rm tol}=10^{-5}$ and $N=256$ at times $T=15$(left) and $T=21$(right).
 All schemes agree to within $L^2$ relative error $2\times10^{-3}$ as reported in Table\,\ref{t:Comp-error}.  The red number on colorbar indicates $\max \{u\}$.}
\label{f:BM2}
\end{figure}

For the critical case the value of $\eta_2$ and $\dcoef$ are tuned to create a strongly pearled interface and a long pearled transient, lasting
roughly from $T=4$ to $T=21$. The bilayer interface pearls transiently, forming 21 pearls, whose  discrete count generates a thresholding effect that
slows the absorption of the dispersed amphiphilic polymer as the interface must generate new pearls to lengthen. During the 21-pearl transient period the pearled bilayer interface undergoes a ``bicycle chain'' meander in which adjacent pearls 
move in opposite directions, either in towards the center or out towards the boundary of the domain, as can be seen in Figure\,\ref{f:BM2} (left). At time $T=21$ the pearls have reduced in size, and two extra pearls form at the points of highest curvature. The formation of the additional pearls facilitates an absorption of mass. As the background level of amphiphilic material is depleted the rate of absorption slows and the the interface returns to an unpearled state, similar to that depicted in  Figure\,\ref{f:BM1} (right) that is able to move freely under a curvature driven motion. No endcap defects are formed in the critical benchmark, and each of the computational schemes are in quantitative agreement.

\subsection{Super-critical benchmark}
\begin{figure}[ht!]
\vspace{-0.1in}
\centering
\includegraphics[width=13.6cm,trim={0.75cm 0.15cm -0.15cm 0.5cm},clip]{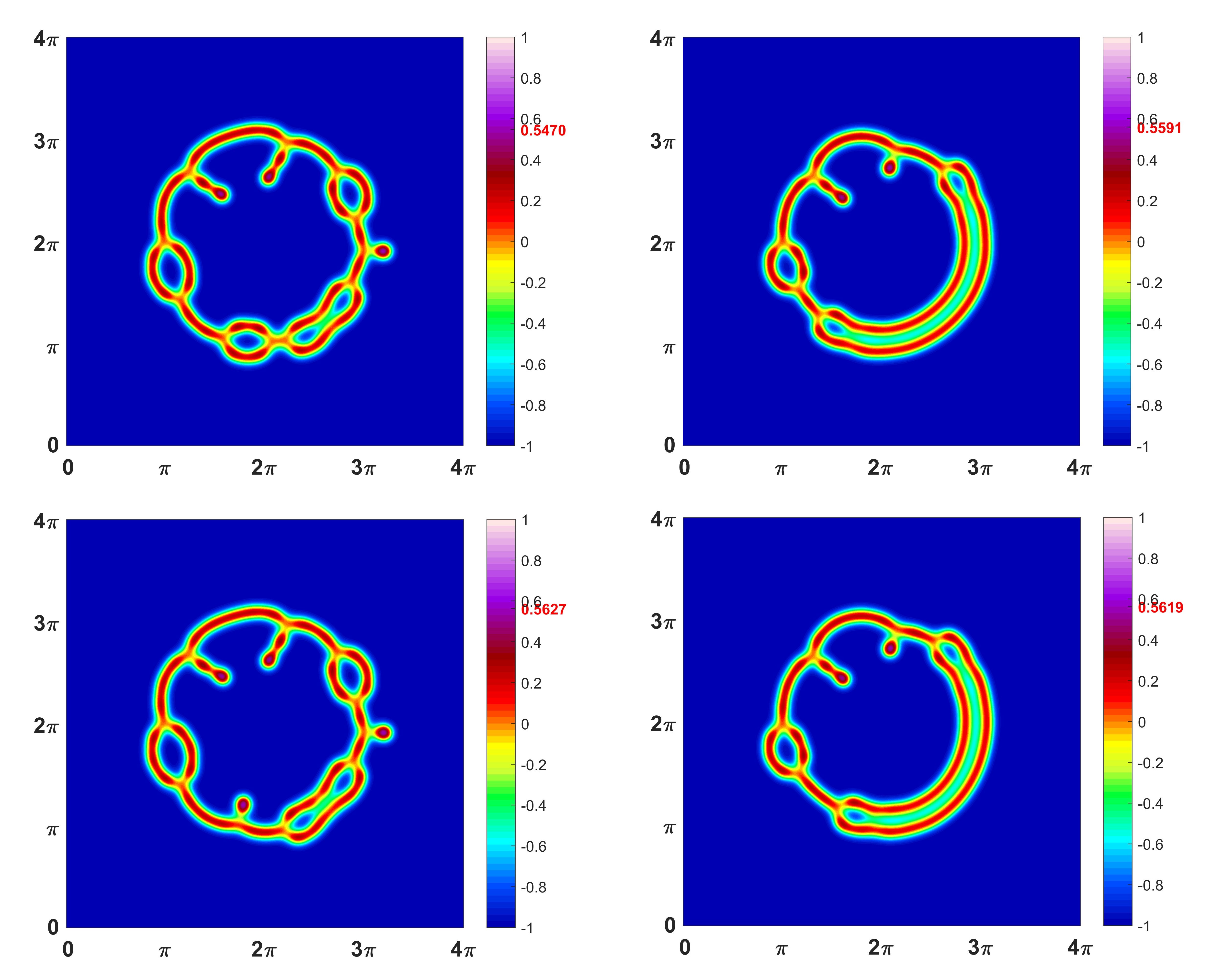}
\vspace{-0.1in}
\caption{Simulations of the super-critical benchmark with $\qtype=0$, $\sigma_{\rm tol} = 10^{-5}$ and $N=256$ at time $T = 50$(left) and $T = 250$(right). 
The top row presents the \PSD simulation and the bottom row represents the \SAV simulation. 
The \IMEX and \SAV simulations are very similar, and the \PSD and \ETD simulations are very similar, but the two groups of simulations disagree, being separated by an $L^2$ relative error of $9\times10^{-2}$, as reported in Table\,\ref{t:Comp-error}.  The red number on colorbar indicates $\max \{u\}$.}
\label{f:BM3}
\end{figure}

\begin{figure}[ht!]
\centering
\vspace{-0.1in}
\includegraphics[width=6.5cm,trim={5cm  8.25cm  4.7cm  8cm},clip]
{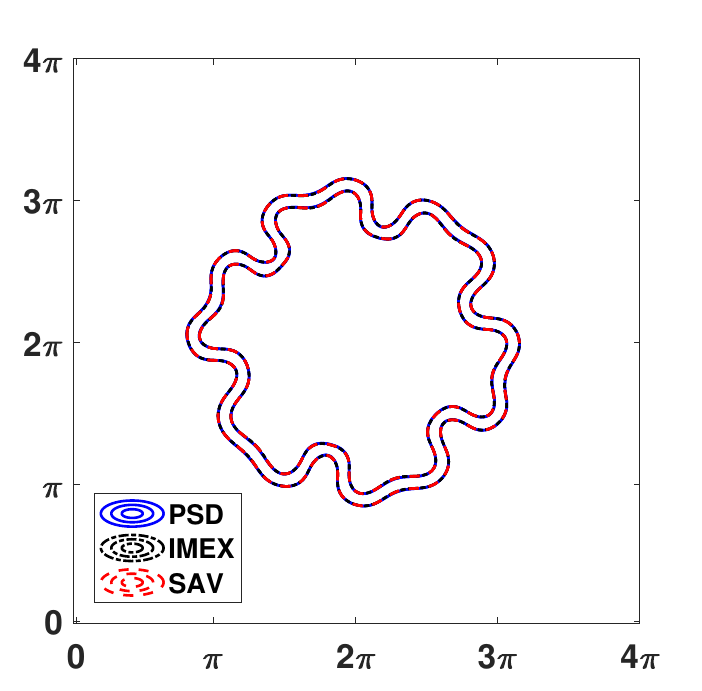}
\includegraphics[width=6.5cm,trim={5cm  8.25cm  4.7cm  8cm},clip]
{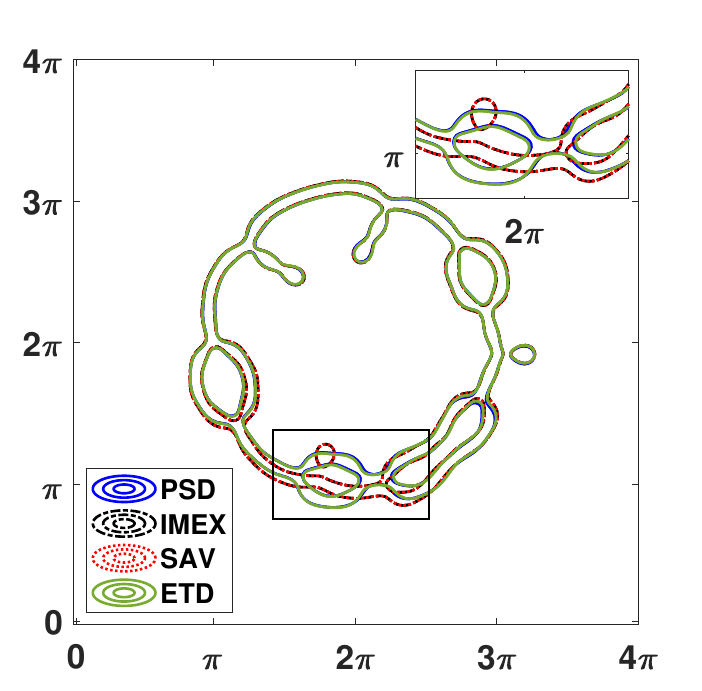}\\
\vspace{-0.1in}
\caption{Contour curves from each of the simulations of each of the schemes with $\sigma_{\rm tol}=10^{-5}$ and $N=256$.  The level set $u=-0.12$  for (left) the sub-critical simulation at $T=10$ and (right) the super-critical benchmark at $T=50$.}
\label{f:contour}
\end{figure}

The sub-critical and super-critical benchmarks differ only in the level of the background material, controlled by the parameter $\dcoef$ in \eqref{dcoef}. The elevated value of this parameter in the super-critical benchmark increases the rate of arrival of mass to the interface, exceeding the interface's capacity to absorb the arriving mass via a curve lengthening flow or by pearl generation. The interface undergoes defect generation. 
For the super-critical benchmark with  $\sigma_{\textrm tol}=10^{-5}$ the output from the four schemes do not agree at leading order, as can be seen in Figure\,\ref{f:BM3}. For the \PSD and \ETD schemes the bilayer interface absorbs material from the background and pearls locally at points of high curvature, and then ejects 8 endcap defects, five of which intersect back with the underlying interface, forming closed loops.  Two of the loops subsequently merge to form an extended loop which grows into a cisternal structure characterized by two long parallel interfaces. The \IMEX and \SAV simulations differ from the \PSD and \ETD, but agree with each-other.  They also produce 8 endcap defects initially, however only four of them subsequently form closed loops. Two of these loops merge, forming a cisternal structure, however there are two small endcaps in the \IMEX and \SAV simulations, in contrast to the one small endcap in the \PSD and \ETD simulation. At longer times the cisternal region grows, consuming structures and at time $T=250$ it leaves one loop,  one long endcap, and one short endcap in all simulations -- 
however in the \SAV and \IMEX simulations the distance between cisternal region and small loop is significantly longer than in \PSD and \ETD simulations. 
Figure\,\ref{f:contour} shows the levels sets corresponding to $u=-0.12$ for the sub-critical and super-critical benchmarks with $\sigma_{\rm tol}=10^{-5}$ and $N=256$, showing their agreement in the sub-critical benchmark and their disparity in the super-critical benchmark.  In the super-critical benchmark the higher rate of absorption driven by the higher initial background level of $u$ produces dynamic choices associated to endcap formation that require greater accuracy than the linearly implicit schemes can achieve at $\sigma_{\rm tol}=10^{-5}$.  If $\sigma_{\rm tol}$ is reduced to $10^{-6},$ then \PSD and \ETD simulations do not change quantitatively, while the \SAV and \IMEX simulations move into quantitative agreement with the \PSD and \ETD schemes. 

\begin{figure}[ht!]
\centering
\vspace{-0.1in}
\includegraphics[width=7cm,trim={2cm  7cm  2.5cm  7cm},clip]{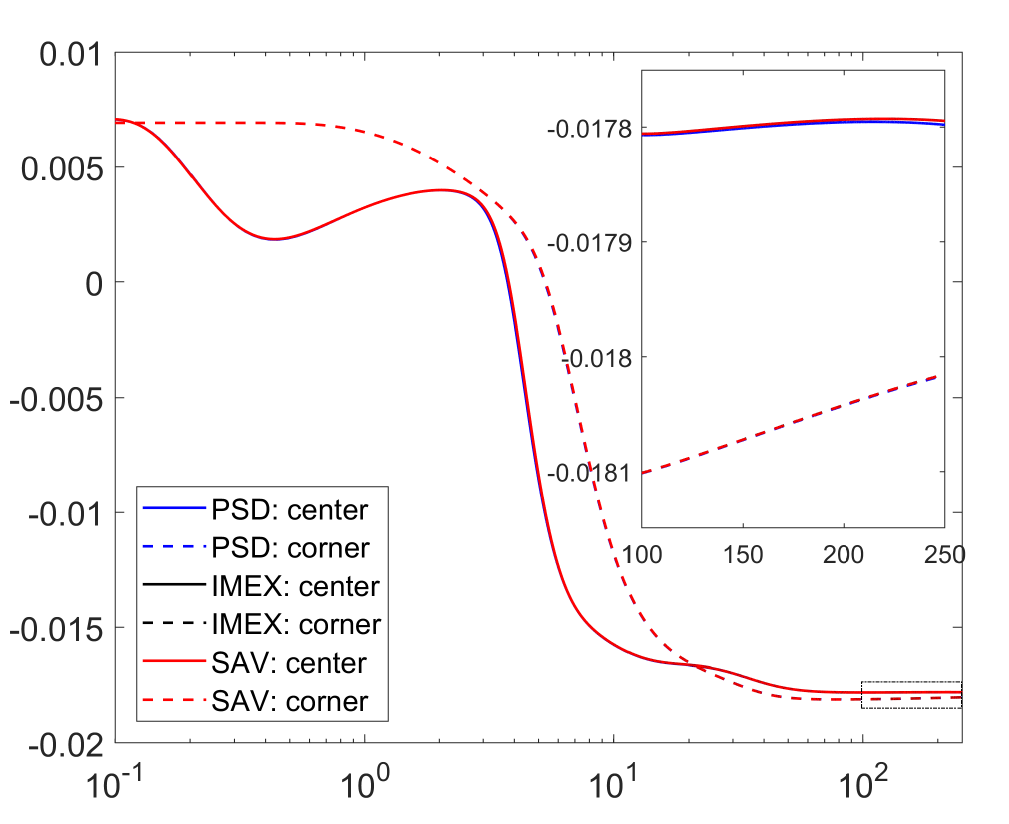}
\includegraphics[width=7cm,trim={2cm  7cm  2.5cm  7cm},clip]{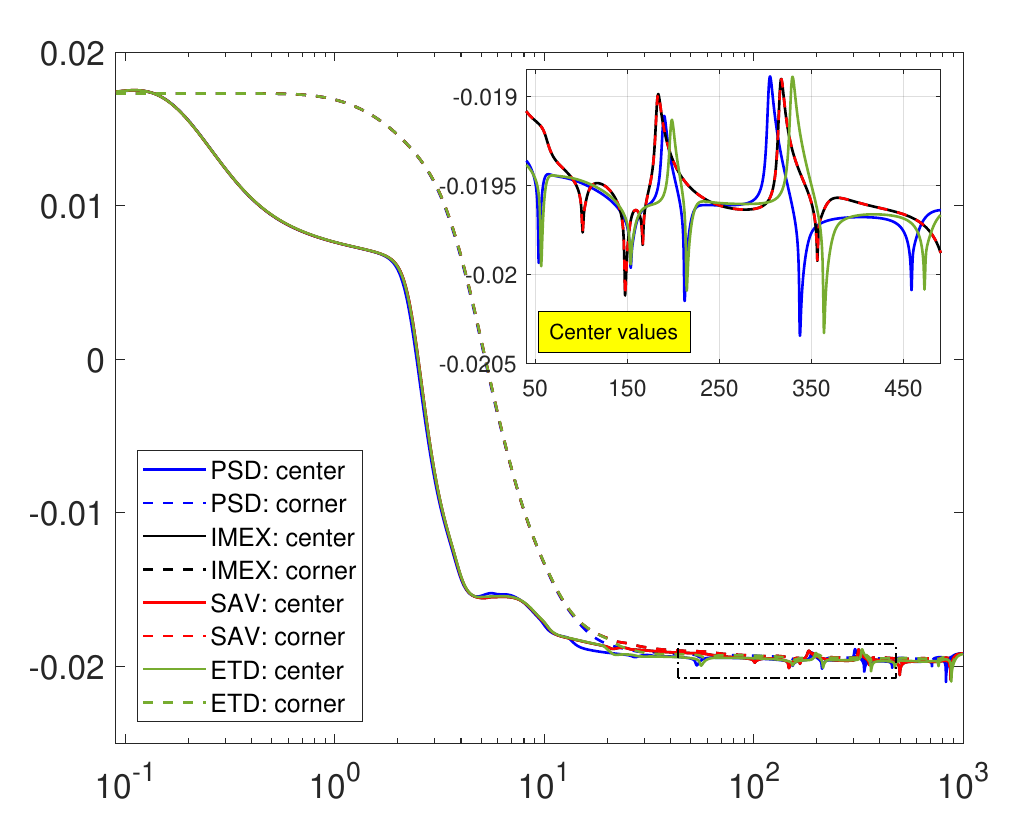}
\vspace{-0.1in}
\caption{Value of $u-b_-$ at center point (solid) and corner point (dashed) of computational domain for the sub-critical (left) and super-critical (right) benchmarks with $\sigma_{\rm tol}=10^{-5}$. Horizontal axis is log of time.}
\label{f:corner-1} 
\end{figure}

The value of $u$ in the far field, away from the interfacial structure, is asymptotically constant at equilibrium and has been shown to be a 
key bifurcation parameter for the onset of pearling, \cite{NP-18, PWu-15}. 
Faithful resolution of this value is essential to an accurate simulation. 
Figure\,\ref{f:corner-1} traces the evolution of the value of $u-b_{-}$ at the domain center (solid lines) and domain corner (dashed lines) for each of the simulation strategies. For the sub-critical simulation no defects are formed and the far-field values of $u$ relax to a tight range of equilibrium values over the time frame $T=75\sim100$.  The super-critical simulations have various defect merging events and each is associated with a small excursion in the background levels. In the inset of Figure\,\ref{f:corner-1} (right) these excursions can be seen at $T=210, 330,$ and $460$ for the \PSD scheme. For the \ETD scheme the excursions are similar but can be delayed by up to $T=20$. Conversely for the \IMEX and \SAV schemes the background levels are in close agreement, recording excursions $T=150, 350,$ and $500,$  but differ in both timing and in number of events from the more accurate \PSD and \ETD simulations.  

\subsection{Foot 1 benchmark}
\begin{figure}[htbp!]
\vspace{-0.1in}
\centering
\includegraphics[width=13.6cm,trim={0.2cm 0.15cm -0.15cm 0.5cm},clip]{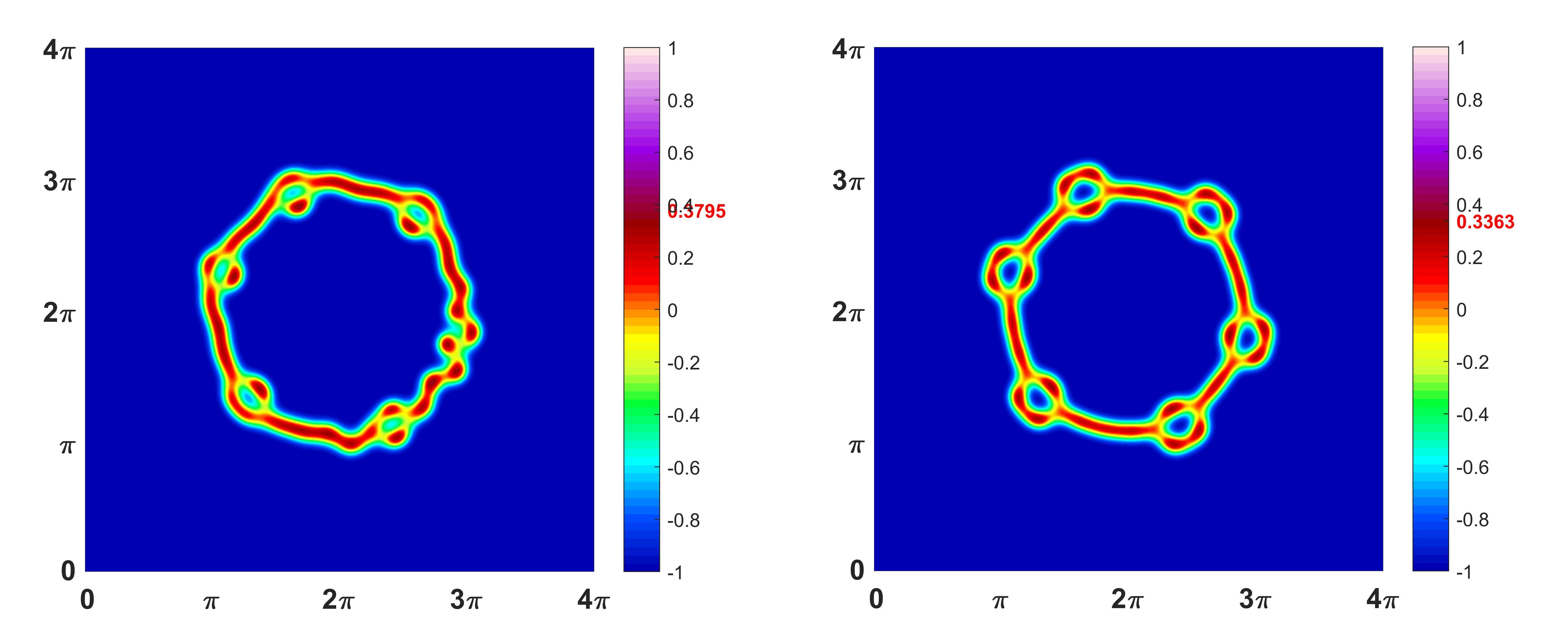}
\vspace{-0.1in}
\caption{Simulation of the Foot 1 benchmark with $\qtype=0.2$, $\sigma_{\rm tol}=10^{-5}$ and $N=256$ at times $T=1.5$(left) and $T=50$(right).
 All three schemes agree to within $L^2$ relative error $3\times10^{-3}$ as reported in Table\,\ref{t:Comp-error}.  The red number on colorbar indicates $\max \{u\}$.}
\label{f:BM4}
\end{figure}

The Foot 1 and sub-critical benchmarks, are identical in initial data and parameters with the exception of the value of the concavity of the well $\Wq$, controlled by the parameter $\qtype$. For Foot 1 we take  $\qtype=0.2$ which increases the value of 
$\alpha_m(\qtype)=\Wq''(b_-)$, as depicted in Figure\,\ref{f:W-qtype}. 
This adjustment raises the energy associated to small, spatially uniform values of $u$, thereby increasing the rate of absorption of material from the bulk. Although the total amount of material in the background is the same in both benchmarks, the increased absorption rate in the Foot 1 benchmark leads to defect formation. We consider only the \PSD\!, \IMEX\!, and \SAV schemes, and each capture these events with quantitative accuracy,  as shown in Table\,\ref{t:Comp-error}. 
In Figure\,\ref{f:BM4} (left) the pearling and defect formation are visible in the lower-right of the bilayer interface already at time $T=1.5$. At time $T=50$ the simulations produce six closed loops place roughly symmetrically around the bilayer interface. This structure is quasi-stable, but eventually evolves onto a double-sheeted bubble similar to that depicted in the right-most panel of the top row of Figure\,\ref{f:BD-numerical}.

\subsection{Foot 2 benchmark}
\begin{figure}[ht!]
\vspace{-0.1in}
\centering
\includegraphics[width=13.6cm,trim={0.75cm 0.1cm -0.15cm 0.5cm},clip]{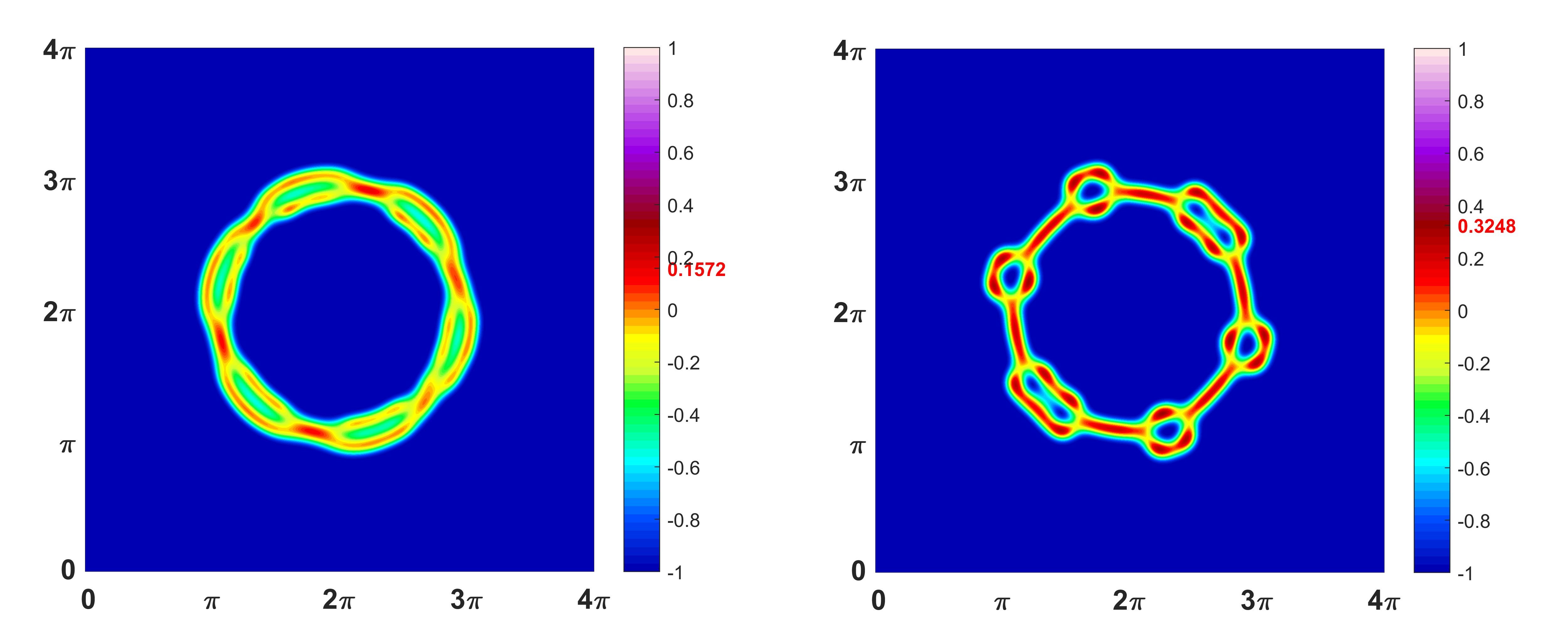}
\vspace{-0.1in}
\caption{Simulation of the Foot 2 benchmark with $\qtype=0.5$, $\sigma_{\rm tol}=10^{-5}$ at times $T=1$ (left) and $T=50$ (right)
for $N=512$.  All three schemes agree to within $L^2$ error $1\times10^{-2}$ as reported in Table\,\ref{t:Comp-error}.  The red number on colorbar indicates $\max \{u\}$.}
\label{f:BM5}
\end{figure}

The Foot 2 and sub-critical benchmarks have an identical setup with the exception of the value of $\qtype$, which is taken to be $\qtype=0.5$ in  Foot 2. This introduces a very strong, nonlinear stiffness, and the large value of $\alpha_m=\Wq''(b_-)\bigl|_{\qtype=0.5}$ significantly increases the energy penalty associated to dispersed amphiphilic material. As a consequence its rate of absorption onto the bilayer interface increases, inducing a curve-splitting bifurcation in which the bilayer interface splits directly in two, as shown in Figure\,\ref{f:BM5} (left) at $T=1$. All three schemes agree qualitatively on the $512\times512$ mesh, producing four loops and two double loops. Grid refinement in Table \,\ref{t:GRE} shows that the $N=256$ grid is insufficient to produce accurate results. Further grid refinement to $N=1024$ yields quantitative agreement with the $N=512$ simulations. The large value of $\Wq''(b_-)$ for $\qtype=0.5$ yields a profile that is much less smooth. The spatial convergence to the far-field value occurs at the exponential rate $\sqrt{\Wq''(b_-)}\Big /\ep$, which is significantly greater for $\qtype=0.5$, necessitating the higher spatial resolution. 

The time-trace of the background levels, $u-b_-$ evaluated at the domain center (solid) and domain corner (dashed), are presented for the Foot 2 benchmark in Figure\,\ref{f:Stepsize} (left). 
It has several notable differences from the sub-critical benchmark presented in Figure\,\ref{f:corner-1} (left).  The most salient distinction is that the large value of  $\alpha_m(0.5)$ greatly increases the temporal rate of absorption of amphiphilic material from the background. 
For the Foot 2 benchmark the background state begins to achieve its equilibrium value at $T=1$ and is fully equilibrated  around $T=7\sim8.$ This is roughly 10-15 times faster than the relaxation for the $\qtype=0$ sub-critical benchmark, depicted in Figure\,\ref{f:corner-1}.

% Table generated by Excel2LaTeX from sheet 'Convergence_Rate'
\begin{table}[htbp]
  \centering
  \caption{$L^2$ grid refinement (absolute) error with the \PSD scheme.}
  \renewcommand{\arraystretch}{1.2}
    \begin{tabular}{|c|c|c|}
    \hline
    \textbf{$N$}     & {256 / 512} & {512 / 1024} \\ \hline
    {Sub-Critical}   & 6.218E-04   &  \\ \hline
    {Critical}       & 3.827E-04   &  \\ \hline
    {Super-Critical} & 2.589e-04   &  \\ \hline
    {Foot 1}         & 8.502E-02   &  \\ \hline
    {Foot 2}         & 1.008       & 5.762E-04 \\ \hline
    \end{tabular}%
  \label{t:GRE}%
\end{table}%

\begin{table}[htbp!]
  \centering
  \caption{$L^2$ temporal convergence errors and rates. The error is determined by comparison to \PSD with a fixed temporal step size $k=10^{-6}$ and $i_{\rm tol}=10^{-11}.$} 
  \renewcommand{\arraystretch}{1.2}
\begin{tabular}{|c|c|c|c|c|c|c|c|c|}
\hline
Schemes  & \multicolumn{2}{c|}{\IMEX} & \multicolumn{2}{c|}{\PSD} & \multicolumn{2}{c|}{\SAV} & \multicolumn{2}{c|}{\ETD} \bigstrut\\ \hline
 fixed $k$ & $L^2$ Error & Rate &  $L^2$ Error & Rate &  $L^2$ Error & Rate &  $L^2$ Error & Rate  \bigstrut\\ \hline
$8\times10^{-2}$ 
& 2.20E-01 &       & 7.03E-05 &       & 2.20E-01 &     &  &      \bigstrut \\ \hline
$4\times10^{-2}$  & 5.38E-02 & 2.03  & 1.77E-05 & 1.99  & 5.38E-02 & 2.03 &  &     \bigstrut\\ \hline
$2\times10^{-2}$  & 1.37E-02 & 1.98  & 4.43E-06 & 2.00  & 1.37E-02 & 1.98 &  &     \bigstrut\\ \hline
$1\times10^{-2}$  & 3.54E-03 & 1.95  & 1.11E-06 & 2.00  & 3.54E-03 & 1.95 &  &     \bigstrut\\ \hline
$5\times10^{-3}$ & 9.31E-04 & 1.93  & 2.79E-07 & 1.99  & 9.31E-04 & 1.93 & 3.47E-01  &     \bigstrut\\ \hline
$2.5\times10^{-3}$ & 2.46E-04 & 1.92  & 8.36E-08 & 1.74  & 2.46E-04 & 1.92 & 1.41E-01  &    1.30     \bigstrut\\ \hline 
$1.25\times10^{-3}$ & 6.47E-05 & 1.93  & 6.89E-08 & 0.28  & 6.46E-05 & 1.93 & 5.05E-02  &    1.48   \bigstrut\\ \hline 
$6.25\times10^{-4}$ & 1.69E-05 & 1.94  & 6.62E-08 & 0.06  & 1.69E-05 & 1.94 & 1.65E-02   &   1.61    \bigstrut\\ \hline 
$3.125\times10^{-4}$ & 4.37E-06 & 1.95  & 5.81E-08 & 0.19  & 4.37E-06 & 1.95 & 5.07E-03   &   1.71 \bigstrut\\ \hline 
$1.563\times10^{-4}$ & 1.12E-06 & 1.96  & %7.58E-08 
& %-0.38 
& 1.12E-06 & 1.96 & 1.48E-03   &   1.77 \bigstrut\\ \hline 
$3.906\times10^{-5}$ & & &  & & & & 1.17E-04  &   1.85 
\bigstrut\\ \hline 
$9.766\times10^{-6}$ & & &  & & & & 8.51E-06   &   1.90 \bigstrut\\ \hline %
$2.441\times10^{-6}$ & & &  & & & & 5.77E-07 & 1.95 \bigstrut\\ \hline 
    \end{tabular}%
  \label{tab:rates}%
\end{table}%

\section{Computational accuracy and efficiency}

The four schemes presented are second order accurate, as verified by the convergence study presented in Table\,\ref{tab:rates}. Nevertheless, the performance of the schemes is not equally accurate nor efficient, particularly as the nonlinear stiffness parameter $\qtype$ is increased. Generally the \ETD scheme requires substantially smaller time steps to achieve competitive local truncation errors. This is consistent with analysis in \cite{JCP-13} which showed that Runge-Kutta based schemes, even fully implicit ones, can lead to larger truncation errors. It is clear that the \ETD scheme achieves second order accuracy, however it incurs a larger constant from amplification of error in the stages due to the presence of large space gradients in the bilayer morphologies. We discuss the relation of accuracy to energy decay, global discretization error, and computational efficiency.

\subsection{Energy decay}
\begin{figure}[ht!]
\centering
\includegraphics[height=5cm,trim={0.5cm  6.5cm  1cm  7.1cm},clip]{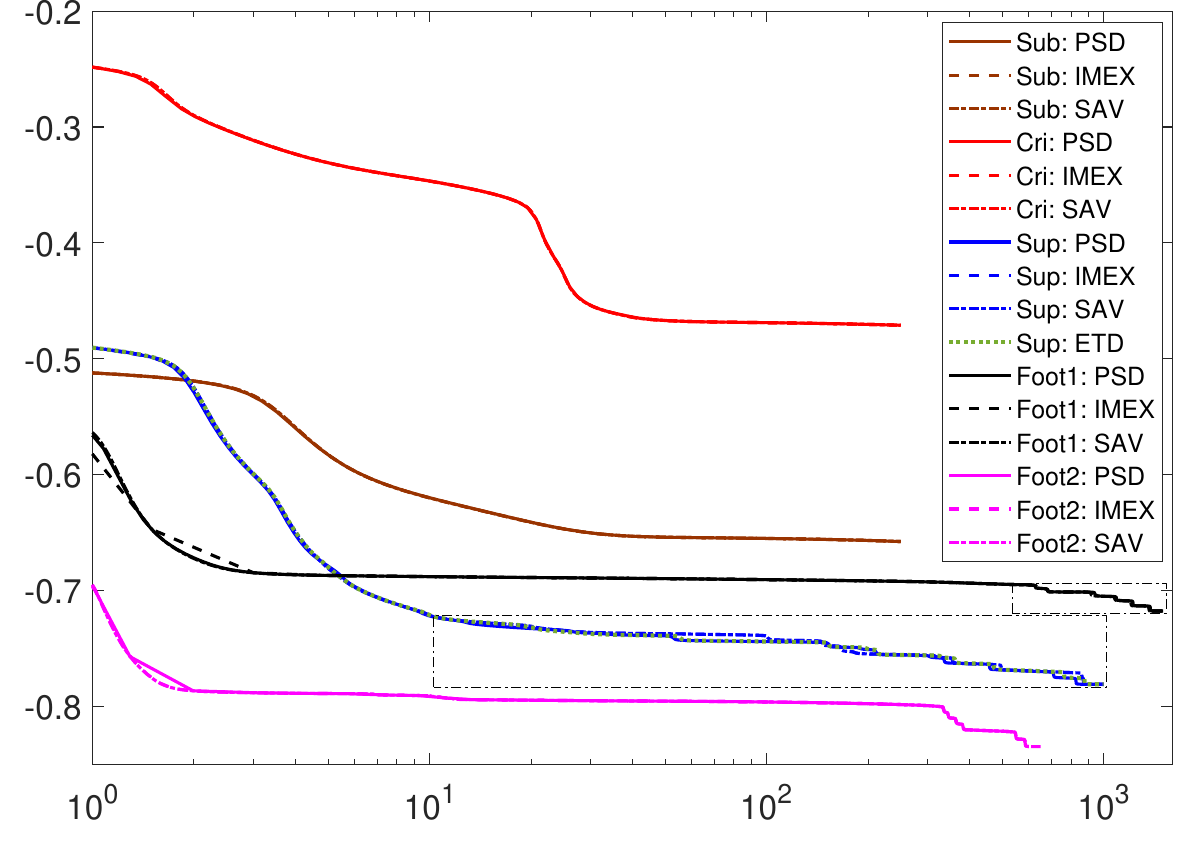}~
\includegraphics[height=5cm,trim={0.5cm  6.5cm  1cm  7.2cm},clip]{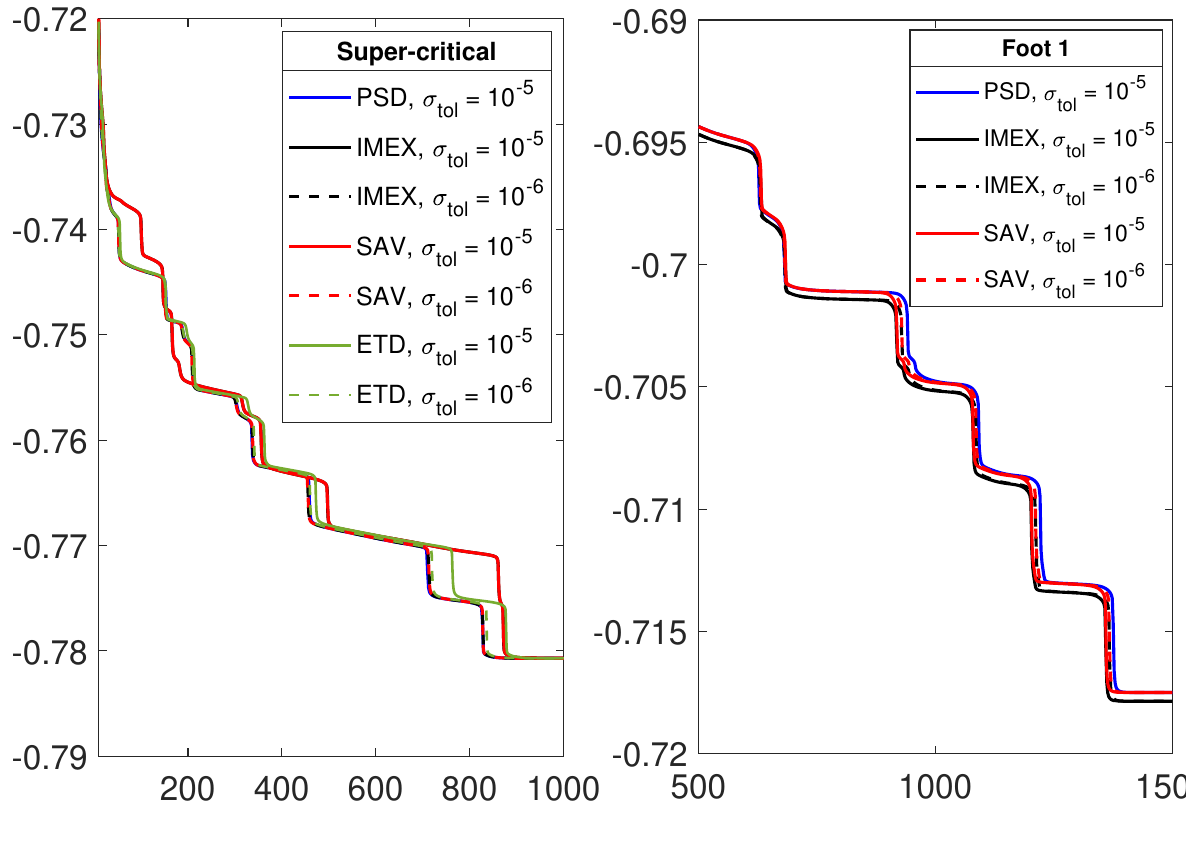}
\vspace{-0.1in}
\caption{(left) System energy verses time on a semilog-x scale for each of the five benchmark problems for each scheme with $\sigma_{\rm tol}=10^{-5}$. The boxed insets for the super-critical (middle) and foot 1 (right) benchmarks show more detail and include results for \IMEX and \SAV with $\sigma_{\rm tol}=10^{-6}$.} 
\label{f:Energy}
\end{figure}

A major feature of gradient schemes is the decay of the overall system energy. Much attention has been given to the construction of
gradient stable schemes for which energy decay is unconditional with respect to the temporal step-size. However in gradient flows that generate a rich variety of structures issues of accuracy move to the forefront and energy decay ideally becomes a consequence of accuracy. 
For the super-critical benchmark, the various competing outcomes are significantly different but have only marginally different energies and considerable accuracy is required for a scheme to differentiate between the available options. As shown in Figure\,\ref{f:Energy} (left), with $\sigma_{\rm tol}=10^{-5}$ for each of the 5 benchmarks the energy decay behavior is very similar and decays uniformly. There are however important differences. As the middle inset shows, for the super-critical benchmark the energy trace for the \IMEX and \SAV simulations are almost indistinguishable from each-other, but diverge from the more accurate \PSD simulation with roughly a 1\% relative error.  The differences in energy decay, and solution $u$, are largely erased for \IMEX and \SAV when $\sigma_{\rm tol}$ is reduced to $10^{-6}$. The \ETD has an energy trace that is more faithful to the \PSD\!, but has a notable excursion for $T\in[750,850]$ that is eliminated for the reduced value $\sigma_{\rm tol}=10^{-6}$. The second inset shows detail of the Foot 1 benchmark. In this case the \PSD\!, \IMEX\!, and \SAV schemes have reasonable quantitative agreement. And error is further reduced by taking $\sigma_{\rm tol}$ to be $10^{-6}$ in \IMEX and \SAV\!\!. These features emphasize that system energy can be a poor proxy for accuracy, and that energy decay is generally a minor benchmark for a gradient flow.

\begin{figure}[ht!]
\centering
\includegraphics[width=7.25cm,trim={2cm  7.25cm  2.5cm  7cm},clip]{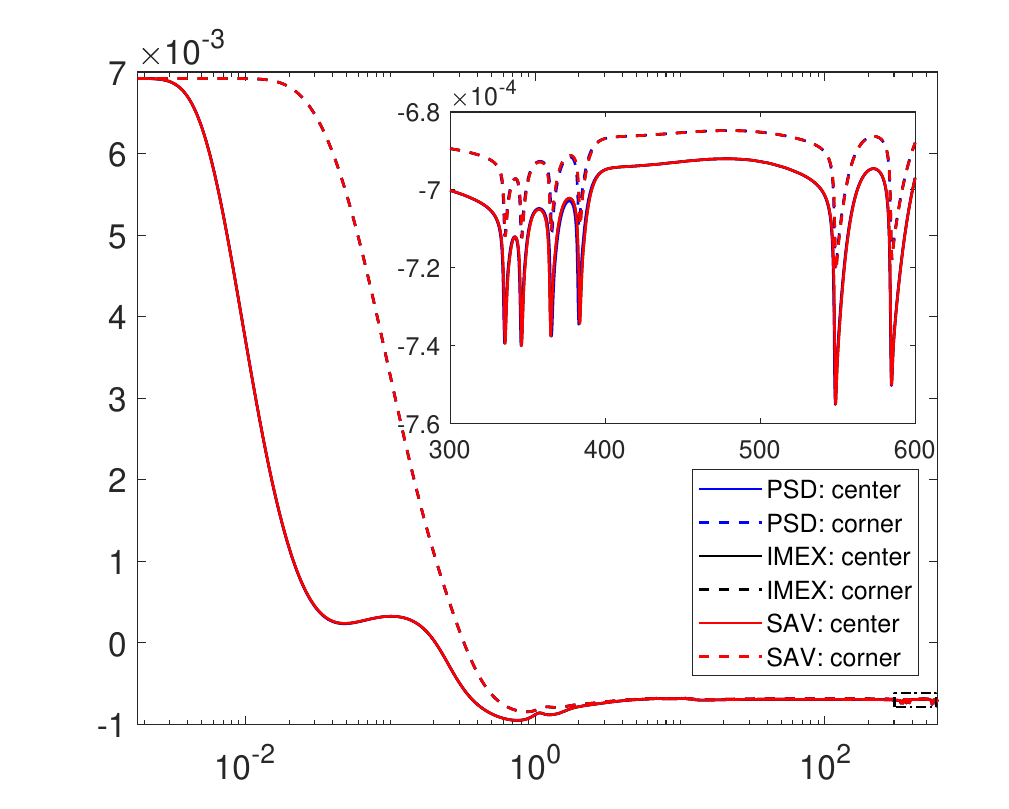}
\includegraphics[width=7.25cm,trim={2cm  7.25cm  2.5cm  7cm},clip]{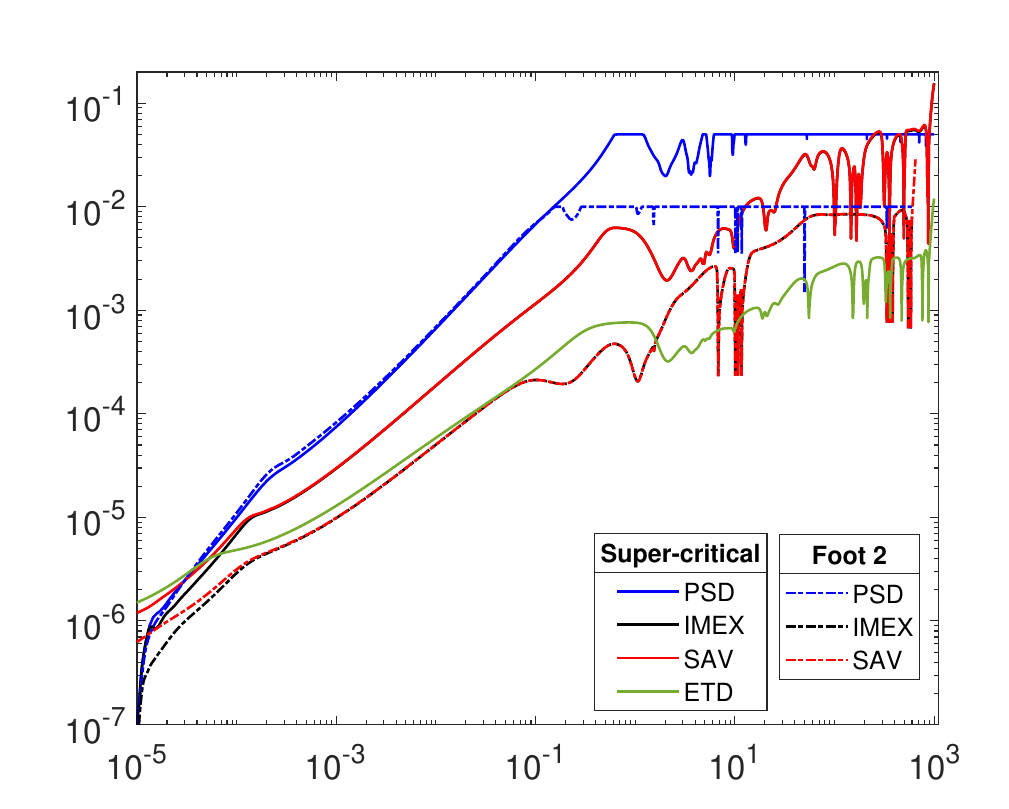}
\vspace{-0.1in}
\caption{(left) Value of $u-b_-$ at center point (solid) and corner point (dashed) of the computational domain for the $\qtype=0.5$ Foot 2 benchmark for $\sigma_{\rm tol}=10^{-5}$ and $N=512.$ (right) Evolution of the adaptive temporal step-size on a log-log scale for each of the four schemes for the $\qtype=0$ super-critical benchmark (solid) and the $\qtype=0.5$ Foot 2 benchmark (dashed). Horizontal axis is log of time.} 
\label{f:Stepsize}
\end{figure}

The time-stepping profiles for the \IMEX and \SAV schemes are remarkably similar, and differ in important ways from that of the \PSD scheme. As shown in Figure\,\ref{f:Stepsize} (right), the \PSD generically takes the largest time step-sizes, and typically hits the maximum step-size ceiling $k_{\rm max}$ shortly after the resolution of the initial transient. This ceiling is required to insure the convergence of the nonlinear iterative scheme and to optimize its performance as measured by FFT per time unit. 
This value is smaller for the stiffer Foot 2 benchmark than for the super-critical benchmark as reported in Table \,\ref{tolPSD}. Indeed the time-step profile for \PSD is largely equivalent for the super-critical and the Foot 2 benchmarks, until it hits the lower value of $k_{\rm max}$ for the Foot 2 benchmark. This is in contrast to the \IMEX and \SAV profiles which are different for the two benchmark problems, but largely agree with each other.  
Each of the schemes has swings in step size of roughly one order of magnitude during the various defect generation and merging events that occur after the initial transient. The step sizes for the \IMEX and \SAV schemes are generically smaller than those for \PSD\!\!, by as much as two orders of magnitude for the stiffer Foot 2 benchmark. However this is offset by the growing number of iterations required for solving the stiffer nonlinear system in this problem. The \ETD scheme has the smallest time steps, typically over an order of magnitude smaller than any of the BDF2 schemes.

An excellent proxy for accuracy is to determine the lowest (critical) value of the background level, as measured by the initial data parameter $\dcoef$ in \eqref{dcoef},
at which a defect is generated within the flow. The onset of a defect is easily detected through the maximum value of $u$, as the maximum value of the bilayer profile in these simulations occurs at $u=0.3566$, while defects and higher codimensional structures such as micelles reside much more deeply in the right well of $\Wq$, with maximum values close to $u=0.74.$  Tracking the temporal evolution of $\max u$ yields a strong dichotomy. 
We fixed the parameters as in the critical benchmark problem but slightly adjusted the value of $\dcoef$ to modify the amount of amphiphilic material in the bulk. The critical $\dcoef$ value, reported in Table\,\ref{f:defect-onsetR} depends upon the local truncation error, but converges to a common value of $\dcoef=0.7526$ with decreasing $\sigma_{\rm tol}.$ Indeed the \PSD scheme is very close to identifying the correct critical value with $\sigma_{\rm tol}=10^{-5}$ while \IMEX and \SAV require a value of $\sigma_{\rm tol}$ of $10^{-7}$ or $10^{-8}$ to achieve similar accuracy.  The time evolution of $\max(u(\cdot,t))$ under \PSD for the critical benchmark parameters and for seven different value of $\dcoef$ is presented in Figure\,\ref{f:defect-onsetL}.

\vspace{0.5cm}
\begin{minipage}{0.95\textwidth}
  \begin{minipage}[c]{0.55\textwidth}
    \centering
    \includegraphics[width=7cm,trim={1.5cm  7.75cm  2cm  8.25cm},clip]{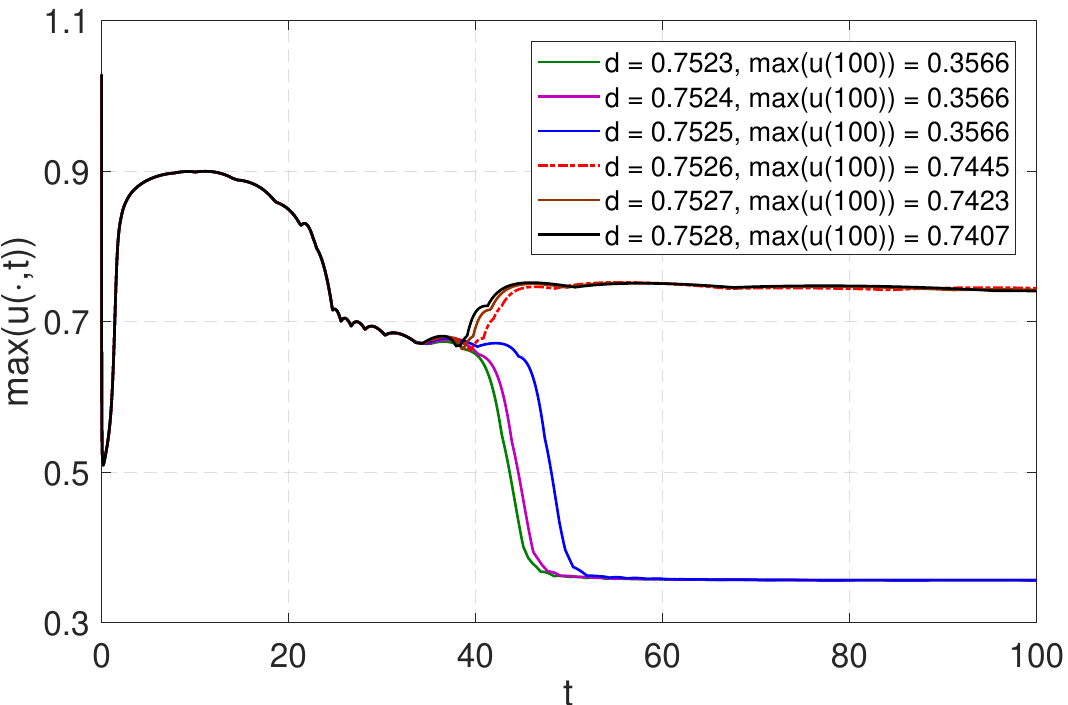}
    \vspace{-0.25cm}
    \captionsetup{type=figure}
    \captionof{figure}{Running value of $\max(u)$ from the \PSD scheme for the critical benchmark problem with $\dcoef=0.7523,...,0.7728$ in increments of $0.0001$ when $\sigma_{\rm tol}=10^{-7}$. When accurately resolved the defect onset occurs at the critical value $\dcoef=0.7526.$}
    \label{f:defect-onsetL}
  \end{minipage}
  \hfill
  \begin{minipage}[c]{0.42\textwidth}
    \centering
    \vspace{-2cm}~\\
    \captionsetup{type=table}
    \captionof{table}{The dependence of the critical value of $\dcoef$ in \eqref{dcoef} upon $\sigma_{\rm tol}$ for each scheme.}
    \vspace{0.25cm}
\begin{tabular}{|c|c|c|c|}
\hline
$\sigma_{\rm tol}$  & \PSD   & \IMEX  & \SAV \bigstrut \\ \hline
$10^{-5}$  & 0.7527 & 0.7540 & 0.7541 \bigstrut \\  \hline
$10^{-6}$  & 0.7525 & 0.7529 & 0.7529 \bigstrut \\  \hline
$10^{-7}$  & 0.7526 & 0.7527 & 0.7527 \bigstrut \\  \hline
$10^{-8}$  & 0.7526 & 0.7526 & 0.7526 \bigstrut \\  \hline
\end{tabular}
\label{f:defect-onsetR}
    \end{minipage}
\end{minipage}
\vspace{0.5cm}

\subsection{Global discretization error verses computational cost}

The definitive measure of accuracy is to compute the global discretization error of a simulation as measured against a known
highly accurate answer. To produce these highly-accurate solutions we conduct a spatial grid refinement study for each benchmark problem and each computational scheme. For all but the stiffest Foot 2 benchmark increasing the grid from $N=256$ to $N=512$ produces consistent results, with solution differences reported in Table\,\ref{t:GRE}. We present results only to the accuracy determined within this grid refinement study. 
Specifically the highly accurate simulations are calculated with the \PSD scheme with $\sigma_{\rm tol}=10^{-9}$ for $\qtype=0$, and with $\sigma_{\rm tol}=3 \times 10^{-8}$ for $\qtype=0.2$. For $\qtype=0.5$, the \IMEX scheme with $\sigma_{\rm tol}=3 \times 10^{-9}$ is used.
The output of these simulations are taken as the highly accurate simulation against which others are compared. For all three schemes, sufficient refinement of $\sigma_{\rm tol}$ lead to a global error that is within the anticipated accuracy of the scheme. Indeed our computations find that a global $L^2(\Omega)$ relative discretization error of $2.5\times10^{-3}$ is generically sufficient to ensure that scheme is quantitatively accurate, with the correct numbers, types, and placements of defects. 

\begin{figure}[ht!]
\centering
\vspace{-.1in}
\includegraphics[width=7cm,height=5.8cm,trim={1cm  7.25cm  1.75cm  7.5cm},clip]{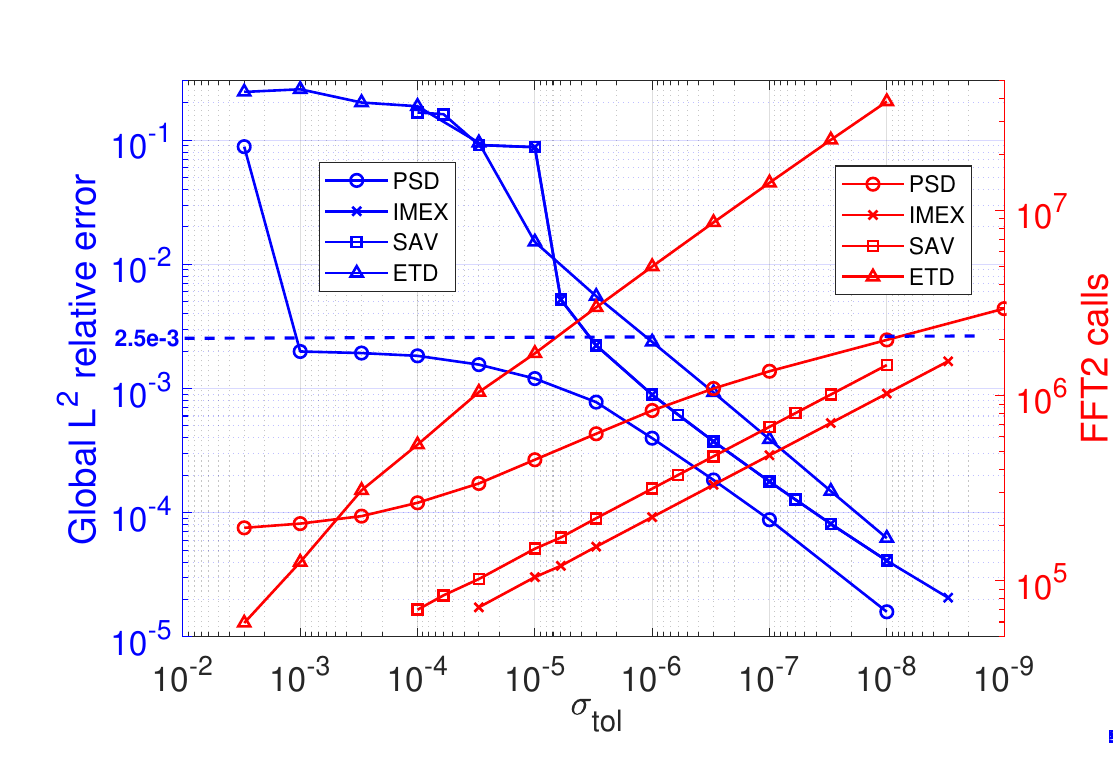}
\includegraphics[width=7.4cm,height=5.8cm,trim={1cm  7.25cm  1cm  7.5cm},clip]{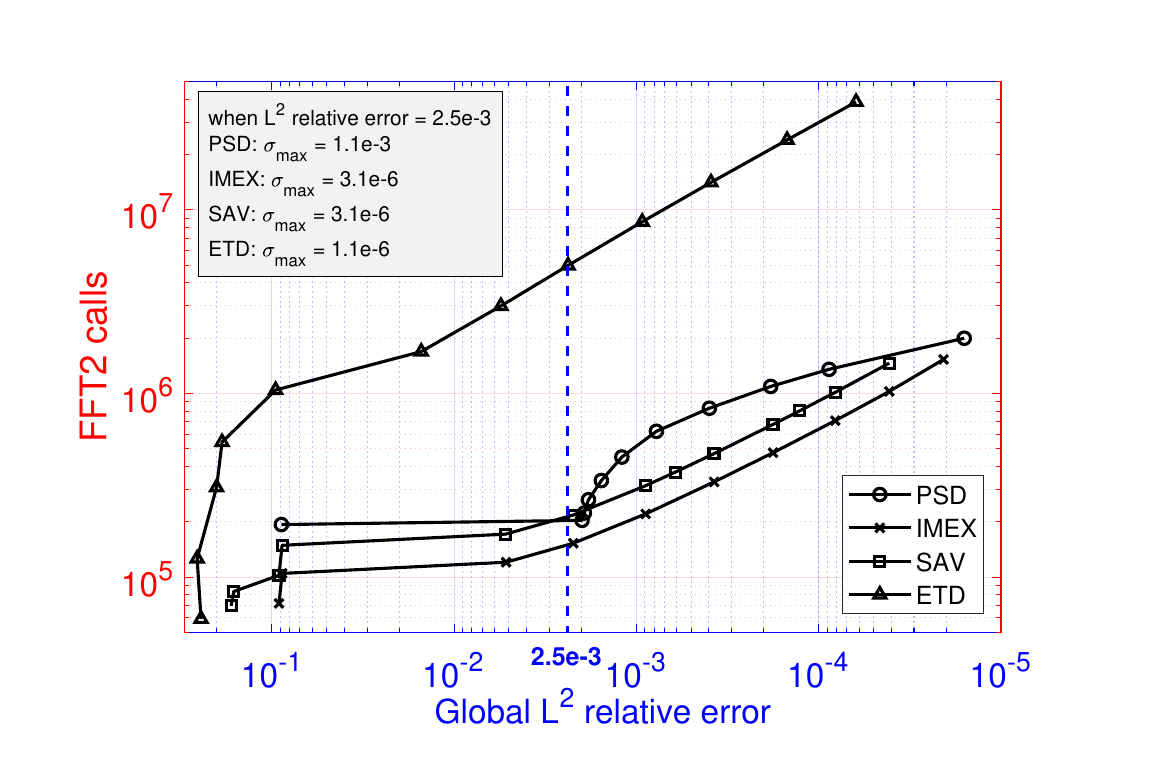}
\vspace{-0.2in}
\caption{(left: blue y-axis and lines) Global $L^2$ relative error verses $\sigma_{\rm tol}$ at $T=250$ for the $\qtype=0$ super-critical benchmark as measured by comparison to the most accurate solution. (left: red y-axis and lines) Computational cost verses  $\sigma_{\rm tol}$ as measured by total number of FFT calls. (right) ``Dollars-per-digit'' or computational cost verses global $L^2$ relative error, plotted parametrically in $\sigma_{\rm tol}.$} 
\label{f:BM3-L2}
\end{figure}

We measure the computational efficiency of the three schemes in two ways. First as global relative truncation error, $G_{\rm rte}$,
verses $\sigma_{\rm tol}$, and then more meaningfully as global discretization error verses FFT calls. This latter is euphemistically referred to as 
the dollars-per-digit metric. The first result, presented in Figure\,\ref{f:BM3-L2} (left), shows the decay in global $L^2$ relative error with decreasing $\sigma_{\rm tol}$.
The blue curves, corresponding to the left (blue) vertical axis, show that all four schemes improve in global accuracy with decreasing  $\sigma_{\rm tol}$. For the super-critical benchmark the linear-implicit 
\IMEX and \SAV schemes are inaccurate for $\sigma_{\rm tol}> 4\times10^{-6}$ and then have global discretization errors that decay linearly on a log-log plot, corresponding to a global discretization error roughly proportional the the $2/3$ power of the local truncation error. The \ETD scheme is more accurate than \IMEX and \SAV for $\sigma_{\rm tol}= 10^{-5}$ but becomes somewhat less accurate than \IMEX and \SAV when decreasing $\sigma_{\rm tol}.$ Conversely, the \PSD is accurate for all $\sigma_{\rm tol}<1\times10^{-3},$ but its global accuracy at first improves sub-linearly with $\sigma_{\rm tol}$ on the log-log scale before setting into the $2/3$ power law relation between global discretization and local truncation errors.  For the linear-implicit schemes the workload as measured by total FFT calls is remarkably linear as function of local truncation error on the log-log curve. Their workload grows approximately as a $-1/2$ power of the local truncation error over three orders of magnitude, with the \IMEX more efficient than the \SAV by a fixed factor of $1.4$ over this range. The \ETD scheme has a significantly higher workload, often by more than an order of magnitude, across all ranges of $\sigma_{\rm tol}.$  The \PSD workload starts out significantly higher than the linear-implicit schemes, but grows more slowly, becoming comparable at very small values of $\sigma_{\rm tol}$. 

A more intuitive comparison of the performance arises from plotting the FFT calls verses the global discretization error, with $\sigma_{\rm tol}$ acting as a parameterization of the curve. This is the dollars-per-digit plot, shown in Figure\,\ref{f:BM3-L2} (right). In this plot, the lowest curve attains the desired global discretization error with the least computation cost. Setting $G_{\rm rte}=2.5\times 10^{-3}$ as an acceptable upper limit, we find that all schemes except \ETD achieve this global tolerance at comparable computational costs that correspond to disparate local truncation errors. The \IMEX scheme is generally the most efficient, hitting the global accuracy mark with $1.5\times10^{5}$ FFT calls at $\sigma_{\rm tol}=3\times 10^{-6}$, while \PSD does so with $2\times 10^5$ FFT calls at a much lower $\sigma_{\rm tol}=10^{-3}$, and \SAV with $2\times10^5$ FFT calls at $\sigma_{\rm tol}=3 \times 10^{-6}.$ 
However the efficiency of the \PSD decays with global relative error above this acceptable upper limit, recovering only at very small global error.   The overall result is a large interval in which the linear-implicit schemes slightly outperform \PSD\!\!. 
The \ETD scheme is not competitive, requiring considerably more computational effort to achieve the same accuracy. A heuristic argument for this result, based upon scaling of trunction error in the thin-interface regime ($\varepsilon\ll1)$ is presented in Appendix B.

\begin{figure}[ht!]
\centering
\vspace{-0.1in}
\includegraphics[width=7cm,height=5.8cm,trim={1cm  7.25cm  2cm  7.5cm},clip]{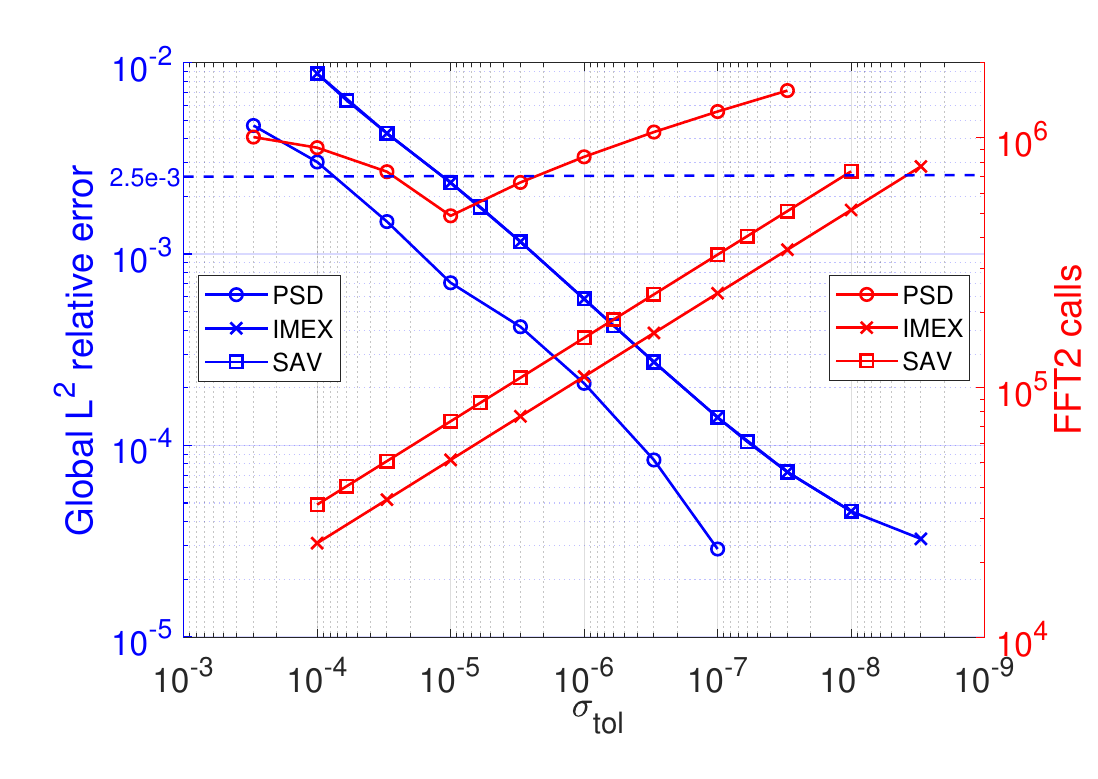}
\includegraphics[width=7.4cm,height=5.8cm,trim={1cm  7.25cm  1cm  7.5cm},clip]{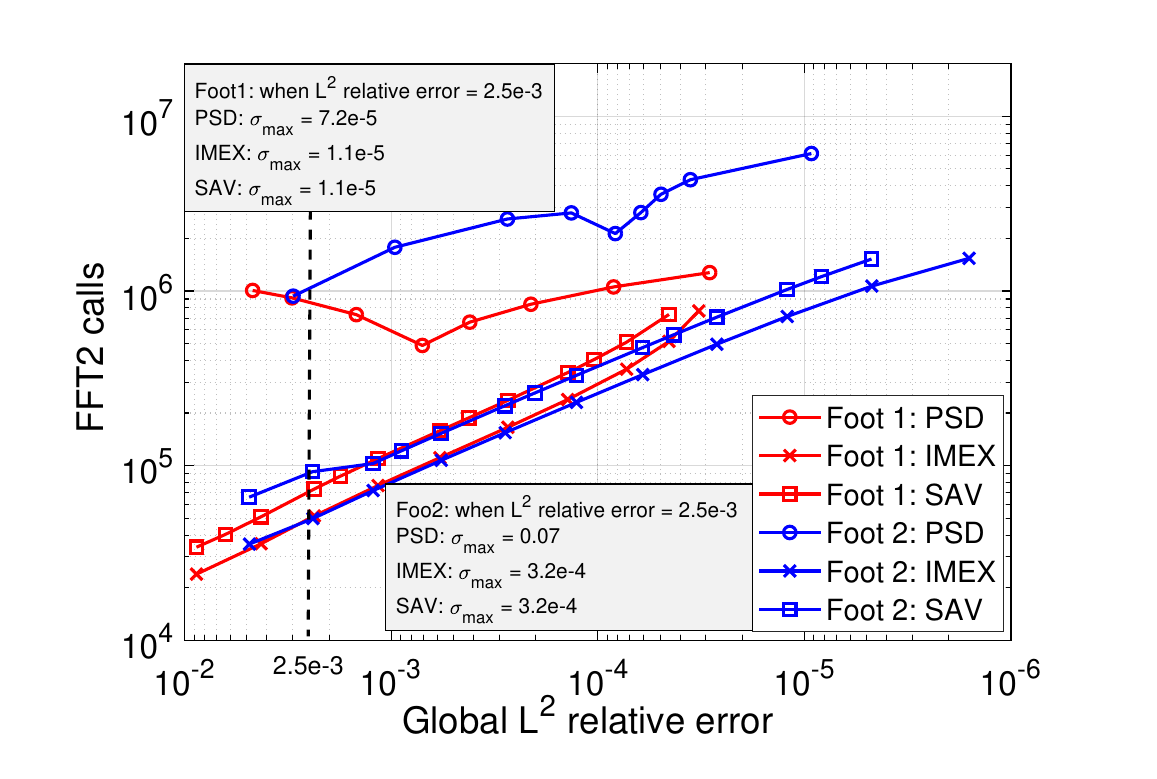}
\vspace{-0.2in}
\caption{(left: blue y-axis and lines) Global $L^2$ relative error verses $\sigma_{\rm tol}$ at $T=50$ for the $\qtype=0.2$ Foot 1 benchmark as measured by comparison to the most accurate solution. (left: red y-axis and lines) Computational cost verses  $\sigma_{\rm tol}$ as measured by total number of FFT calls. (right) ``Dollars-per-digit” or computational cost verses global $L^2$ relative error for Foot 1 and Foot 2 benchmarks, plotted parametrically in $\sigma_{\rm tol}$. } 
\label{f:BM4-L2}
\end{figure}

For the stiffer Foot 1 benchmark simulations with  $\qtype=0.2$ the linear-implicit schemes perform at a similar level to the $\qtype=0$ benchmarks, while the nonlinearly implicit \PSD experiences slower convergence in its nonlinear solver. As shown in Figure\,\ref{f:BM4-L2} (left), the global error for each scheme is an approximately linear function of local truncation error on the log-log scale, corresponding to a power law exponent in the range $0.5\sim0.6$ that is slightly reduced from the $2/3$ exponent observed for the super-critical benchmark. 
The computational efficiency plot, Figure\,\ref{f:BM4-L2} (right), the data for both Foot 1 and Foot 2 benchmarks are compared. The linear-implicit schemes substantially outperform the nonlinear-implicit \PSD\!, 
with the \IMEX scheme preserving its proportional efficiency  over \SAV over two orders of magnitude of global discretization error. 
For the linear-implicit schemes the computational cost is very similar for Foot 1 and 2, with the Foot 2 simulations slightly more accurate due to the increase in spatial resolution to $N=512.$ 
Conversely, the nonlinear-implicit \PSD requires significantly more effort with increasing $\qtype$ as the iteration count in the nonlinear solver increases significantly. The minimal cost for \SAV to achieve the acceptable global discretization error is roughly 1.4 that of \IMEX\!\!.
It is worth noting that \PSD is comparably more efficient at lower global error; indeed it requires only 5 and 20 times the computational effort of \IMEX to achieve an error of $7\times10^{-4}$ for the Foot 1 and Foot 2 benchmarks, respectively.

\begin{figure}[ht!]
\vspace{-0.1in}
\centering
\includegraphics[width=12cm,trim={0cm 9.25cm 0cm 9cm},clip]{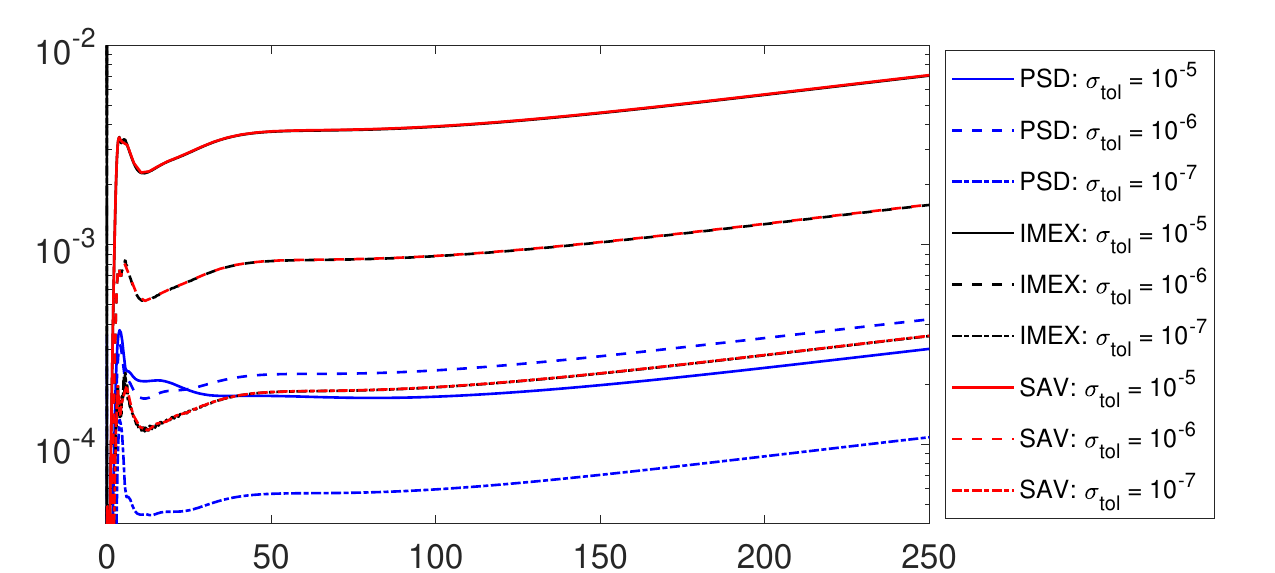}\\
\includegraphics[width=12cm,trim={0cm 9.25cm 0cm 9cm},clip]{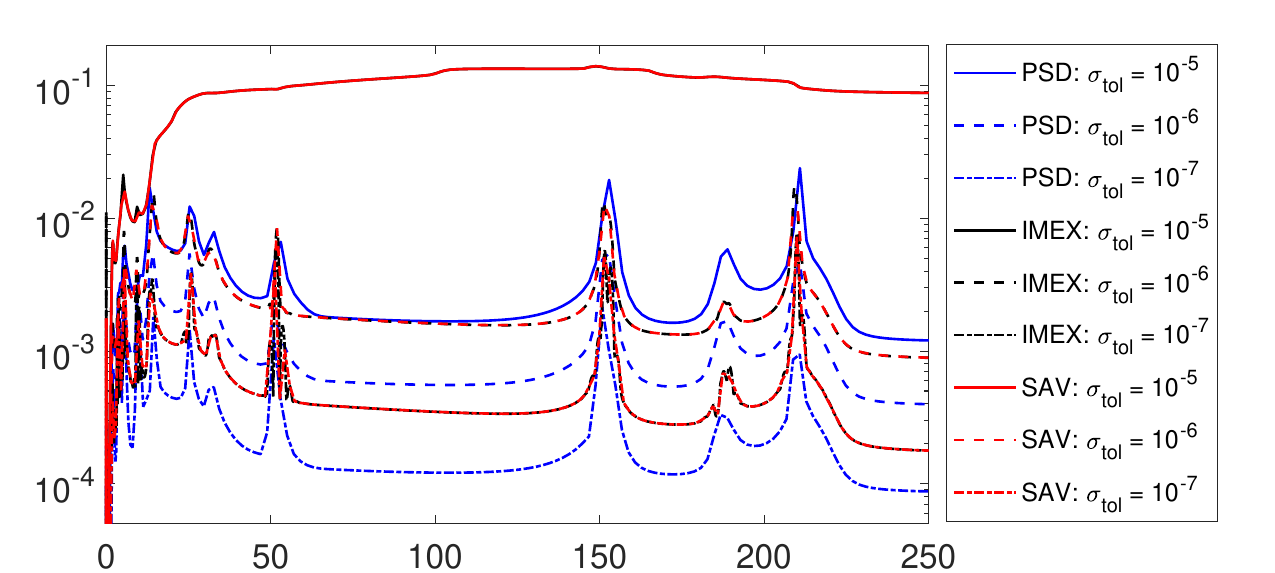}
\vspace{-0.1in}
\caption{Time evolution of the global $L^2$ relative error between output of the three schemes and the highly accurate solution for $\sigma_{\rm tol}=10^{-5}, 10^{-6},$ and $10^{-7}$ for (top) the sub-critical benchmark and (bottom) the super-critical benchmark.}
\label{f:L2temp_error}
\end{figure}

 In Figure\,\ref{f:L2temp_error} the temporal trace of the global error is plotted for local truncation errors of $\sigma_{\rm tol}=10^{-5}, 10^{-6},$ and $10^{-7}$. In all cases the \PSD is the most accurate, generically by an order of magnitude at the same local truncation error. However the accuracy for \PSD increases only modestly with decreasing $\sigma_{\rm tol}$ while \SAV and \IMEX schemes have more significant improvements. 
 For the sub-critical benchmark the global error accumulates  slowly in each of the schemes as the shape of the interface evolves and inaccuracies in its location accumulate. 
 For the super-critical benchmark the error has peaks at each of the major defect merging events that occur at $t=50, 150, 185, 210.$ These peaks reflect the impact of slight timing errors in the defect merging events and in the spatial structure of the merging transient. Each scheme shows about a half-order of magnitude loss of accuracy during the merging that is recovered afterwards. This holds except for the \SAV and \IMEX schemes with $\sigma_{\rm tol}=10^{-5}$ which are both insufficiently accurate to capture the correct sequencing of the defect evolution.  
 
\section{Conclusion}

We have demonstrated that the morphological complexity that develops within the gradient flows of the FCH energy requires accuracy for faithful representation. The benchmark problems place a complex labyrinth of saddle points between the initial data and the end state solution. The saddle points' energies differ by algebraically small orders of $\ep\ll 1.$ Unlike problems in one space dimension which manifest exponentially long residence times, \cite{JCP-13}, resolving the algebraically small differences in the energy landscape makes these benchmarks ideal: simple to code, quick to simulate, and effective at exposing the trade-offs between accuracy and efficiency in a stiff, highly non-convex problem.

These benchmarks model the chemical and material science problems for which computational accuracy is crucial. Small errors in the resolution of the structure of different configurations generate divergent alternate temporal evolutions and errors that grow to become leading order. The impact of this is magnified as the nonlinear stiffness in the model is increased. The nonlinear solve requires in the more strongly implicit \PSD approach tends to raise the overall accuracy of the scheme, and for less-stiff forms of the model this compensates for the increased computational effort required for the iterative solver. The result is that the linear-implicit and nonlinear-implicit models are comparable. However for the more nonlinearly stiff versions of the model, the linear-implicit schemes require no tuning and experience only modest decline in efficiency, while the nonlinear-implicit \PSD requires tuning of the error tolerance and maximum time-step parameters to optimize its performance. Despite this tuning the efficiency of the \PSD scheme falls behind the linear-implicit schemes by a factor that is comparable to the increase in stiffness, as measured by the left-well concavity $\alpha_m(\qtype)=\Wq''(b_-).$ 

Within the linear-implicit schemes the performance of the \IMEX and \SAV schemes are almost indistinguishable. Their global accuracy as a function of local truncation error are almost identical. The only discrepancy lies in the computational effort which is routinely a factor of 1.4 larger for the \SAV scheme. This is a result of the extra steps required to resolve the larger \SAV system of equations. Beyond the guarantee of the decay of the associated modified energy, it is difficult to identify a feature in the \SAV scheme in which it improves upon the simpler \IMEX approach. Far and away the most important step in balancing the linear-implicit schemes is selecting a proper linear term for the implicit step. 
Given the theoretical understanding of the importance of the background state (the value of $u$ away from non-trivial structures), it is reasonable and efficient to use the linearization about the spatially constant state $u\equiv b_-$. 
We generalize this to the family of operators presented in \eqref{e:SAV-L0}, and find that the choice of $\beta_1+\beta_2\approx 3$ provides optimal performance, with the choice $\beta_1=2$ and $\beta_2=1$ corresponding to the linearization about the spatially constant background state. These constant coefficient linear operators are trivially inverted in the spatially periodic setting considered herein.  It certainly may not be the case that such a convenient and efficient linear-implicit operator is available in all systems. The \ETD scheme does not seem to have competitive accuracy in the thin interface regime of the FCH system. The \ETD formulation has been proven effective at handling linear stiffness. It places the higher-order differential operators into a semi-group where they are more stable to discretization error. However, as argued in Appendix B, the local truncation error in the \ETD scheme seems to have poorer scaling with respect to interfacial thickness in the thin interface regime $\varepsilon\ll1$ than \IMEX type schemes.   The large spatial gradients presented by the bilayer interfaces in the super-critical benchmark problem lead to an amplification of error in the \ETD Runge-Kutta approximation.

\begin{figure}[ht!]
\vspace{0.1in}
\centering
\includegraphics[width=14.5cm]{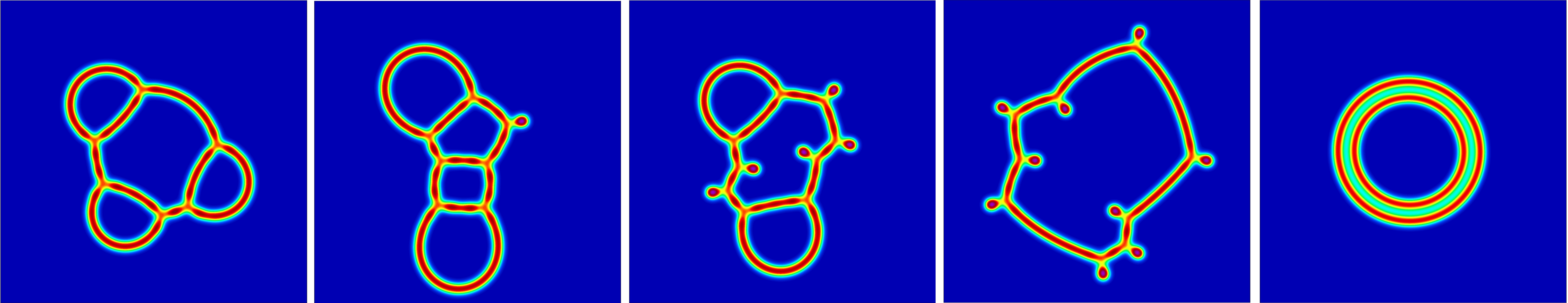}\\
\includegraphics[width=14.5cm]{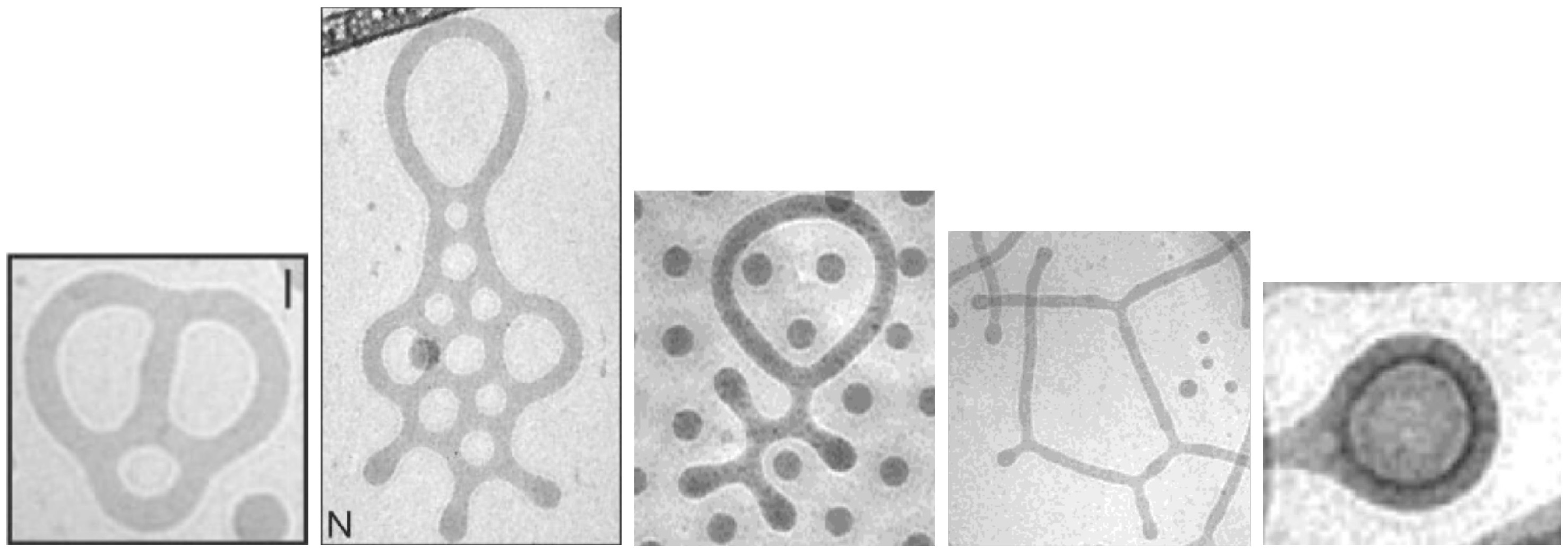}
\vspace{-0.05in}
\caption{(top-row) Approximate equilibrium states computed from super-critical benchmark initial data and parameters except for values of $\eta_2$ taken as $2.55\varepsilon, 2.6\varepsilon, 2.65\varepsilon, 2.8\varepsilon, 2.85\varepsilon$ (left-to-right). Final times are $T=5K, 50K, 100K, 50K, 3K$ respectively. (bottom-row) Experimental comparisons showing (left-to-right) bubbles, bubbles with endcaps, bubbles and branched endcaps, long-branched filaments with endcaps, and double-sheeted bubbles (bubble inside of bubble). Figures 3I and 3N from \cite{Bates-BD}, reprinted with Permission from the AAAS. Figures 5A, 5B, and 9C. Reprinted (adapted) with permission from  \cite{Bates-04}. Copyright (2004) American Chemical Society.}
\label{f:BD-numerical}
\end{figure}

As a final demonstration of the complexity possible within the FCH gradient flow, we present a series of computations that show a putative equilibrium state resulting from the gradient flow of the initial data from the super-critical benchmark, see Figure\,\ref{f:BD-numerical}. The only variation is in the value of the parameter $\eta_2$, which represents the aspect ratio of the amphiphilic molecule. The decreasing values of $\eta_2$ correspond to the increasing values of $w_{\rm PEO}$ in the horizontal axis of the experimental bifurcation diagram presented in Figure\,\ref{f:Bates}. Perhaps the fundamental contribution of this numerical study lies in the suggestion that the shapes of the final structures produced in these casting problems are not uniquely determined by the properties and densities of the molecules they are composed of, but also depend upon the history of the morphology. Once defects are induced by transient flow, they become an integral part of the energy landscape and can entrap the gradient flow at a rich variety of local minima. These gradient flow transients form an intriguing phylogenesis, whose resolution requires significant accuracy. 

%%%% Acknowledgments %%%%%%%%
\section*{Acknowledgements}
A.~Christlieb acknowledges support from NSF grant DMS-1912183. 
K.~Promislow acknowledges support from NSF grant DMS-1813203 and DMS-2205553. 
Z.~Tan recognizes support from the China Scholarship Council under 201906160032.
B.~Wetton recognizes support from a Canadian NSERC grant.  
S.M.~Wise recognizes support from NSF grants DMS-2012634 and DMS-2309547. 

\section*{\large Appendix A. Proof of the energy decay in \SAV -- Theorem\,\ref{t:SAV} }
\label{appendix:A}
From the relations \eqref{4.3}-\eqref{4.4} the \SAV scheme with fixed time-step $k>0$ takes the form
\begin{align}  \label{2.1}
 \frac{3u^{n+1}\!-\!4u^n\!+\! u^{n-1}}{2k} \!&=\! \Delta\mu^{n+1},  \\[1 \jot]
\label{2.2}
 \mu^{n+1}\!&=\! \mathcal{L}_0u^{n+1}\!+\! \mathcal{L}_1 \bar{u}^{n+1}\!+\! r^{n+1} X^{n+1}, \\[1 \jot]
\label{2.3}
3r^{n+1}-4r^n + r^{n-1} &=  \int_{\Omega} \frac{1}{2} X^{n+1} (3u^{n+1}-4u^n+u^{n-1}) \text{d}x,
\end{align}
where $\bar{u}^{n+1} = 2u^n-u^{n-1}$ and $X^{n+1} = \frac{V[\bar{u}^{n+1}]}{\sqrt{\mathcal{E}_1[\bar{u}^{n+1}]+D_0}} $.
Taking the $L^2$ inner product of \eqref{2.1} with $\mu^{n+1}$, and \eqref{2.2} with $3u^{n+1}\!-\!4u^n\!+\! u^{n-1}$, we have
\begin{align}   \label{2.5}
-2k\|\nabla\mu^{n+1}\|^2 
&= (3u^{n+1}\!-\!4u^n\!+\! u^{n-1}, \mu^{n+1}) \nonumber \\
&= (\mathcal{L}_0 u^{n+1}, 3u^{n+1}\!-\!4u^n\!+\! u^{n-1}) \nonumber + (\mathcal{L}_1 \bar{u}^{n+1}, 3u^{n+1}\!-\!4u^n\!+\! u^{n-1})  \nonumber  \\
&~~~~ +  (r^{n+1} X^{n+1},3u^{n+1}\!-\!4u^n\!+\! u^{n-1})  \nonumber  \\[1 \jot]
&=:I_1 + I_2 + I_3.
\end{align}

From the identity 
\begin{align}   \label{2.6}
2a(3a-4b+c) = \left( |a|^2 + |2a-b|^2\right) - \left( |b|^2 + |2b-c|^2\right) + |a-2b+c|^2,
\end{align}
we deduce that
\begin{align}   \label{2.7}
I_1 =\;&\frac{1}{2}\left(\mathcal{L}_0u^{n+1},u^{n+1}\right)+\frac{1}{2}\left(\mathcal{L}_0(2u^{n+1}-u^{n}),2u^{n+1}-u^{n} \right) \nonumber \\
&-\frac{1}{2}\left(\mathcal{L}_0u^{n},u^{n} \right)-\frac{1}{2}\left(\mathcal{L}_0(2u^{n}-u^{n-1}),2u^{n}-u^{n-1} \right)  \nonumber \\[1 \jot]
& + \frac{1}{2}\left(\mathcal{L}_0(u^{n+1}-2u^{n}+u^{n-1}),u^{n+1}-2u^{n}+u^{n-1} \right).
\end{align}

From the identity
\begin{align}    \label{2.8}
2(2b\!-\!c)(3a\!-\!4b\!+\!c) =\left(|a|^2 \!+\! |2a\!-\!b|^2\!-\!2|a\!-\!b|^2\right)\!-\!\left(|b|^2\!+\!|2b\!-\!c|^2\!-\!2|b\!-\!c|^2 \right) -3|a-2b+c|^2,
\end{align}
we rewrite $I_2$ as
\begin{align}  \label{2.9}
I_2 = \; & \frac{1}{2}\left(\mathcal{L}_1u^{n+1},u^{n+1}\right)+\frac{1}{2}\left(\mathcal{L}_1(2u^{n+1}-u^{n}),2u^{n+1}-u^{n} \right)- \left(\mathcal{L}_1(u^{n+1}-u^n),u^{n+1}-u^n\right)  \nonumber \\[1 \jot]
&  -\frac{1}{2}\left(\mathcal{L}_1u^{n},u^{n} \right) -\frac{1}{2}\left(\mathcal{L}_1(2u^{n}-u^{n-1}),2u^{n}-u^{n-1} \right) +  \left(\mathcal{L}_1(u^{n}-u^{n-1}),u^{n}-u^{n-1}\right)  \nonumber \\[1 \jot]
& +\frac{3}{2} \left(-\mathcal{L}_1(u^{n+1}-2u^{n}+u^{n-1}),u^{n+1}-2u^{n}+u^{n-1} \right).
\end{align}

Multiplying \eqref{2.3} by $2r^{n+1}$ and using identity \eqref{2.6}, we get
\begin{align}  \label{2.10}
I_3 %= 2r^{n+1} \left( 3r^{n+1}-4r^n + r^{n-1} \right) 
= \left( |r^{n+1}|^2 + |2r^{n+1}-r^n|^2 \right) - \left( |r^{n}|^2 + |2r^{n}-r^{n-1}|^2 \right) + \big |r^{n+1}-2r^{n}+r^{n-1} \big |^2.
\end{align}

Finally, since $\mathcal{L}_{\text{\SAV\!\!}}:=\mathcal{L}_0+\mathcal{L}_1$ we may combine \eqref{2.5}, \eqref{2.7}, \eqref{2.9} and \eqref{2.10} to deduce
\begin{align}  \label{2.11}
0 \geq & -2k\|\nabla\mu^{n+1}\|^2 \nonumber \\[1 \jot]
= & ~
\frac{1}{2}\left(\mathcal{L}_{\text{\SAV}}u^{n+1},u^{n+1}\right)
+\frac{1}{2}\left(\mathcal{L}_{\text{\SAV}}(2u^{n+1}-u^{n}),2u^{n+1}-u^{n} \right)
+ |r^{n+1}|^2 +  |2r^{n+1}-r^n|^2 
\nonumber \\[1 \jot]
%&  \nonumber \\[1 \jot]
& -\frac{1}{2}\big(\mathcal{L}_{\text{\SAV}}u^{n},u^{n} \big)
-\frac{1}{2}\left(\mathcal{L}_{\text{SAV}}(2u^{n}-u^{n-1}),2u^{n}-u^{n-1} \right) 
- |r^{n}|^2 -  |2r^{n}-r^{n-1}|^2
\nonumber \\[1 \jot]
&- \left(\mathcal{L}_1(u^{n+1}\!-\!u^n),u^{n+1}\!-\!u^n\right) 
+ \left(\mathcal{L}_1(u^{n}\!-\!u^{n-1}),u^{n}\!-\!u^{n-1}\right) 
 \nonumber \\[1 \jot]
& + \frac{1}{2}\left(\mathcal{L}_0(u^{n+1}-2u^{n}+u^{n-1}),u^{n+1}-2u^{n}+u^{n-1} \right) \nonumber \\[1 \jot]
&+ \frac{3}{2} \left(-\mathcal{L}_1(u^{n+1}-2u^{n}+u^{n-1}),u^{n+1}-2u^{n}+u^{n-1} \right)  
%\nonumber \\[1 \jot]& 
+ |r^{n+1}-\!2r^{n}+ r^{n-1}|^2. \nonumber \\[1 \jot]
= & ~ \cE_{\rm aux}\big(u^{n+1},u^{n},r^{n+1},r^{n}\big) - 
\cE_{\rm aux}\big(u^{n},u^{n-1},r^{n},r^{n-1}\big) + |r^{n+1}-\!2r^{n}+ r^{n-1}|^2 \nonumber \\[1 \jot]
& + \frac{1}{2}\left(\mathcal{L}_0(u^{n+1}-2u^{n}+u^{n-1}),u^{n+1}-2u^{n}+u^{n-1} \right)
\nonumber \\[1 \jot]
& + \frac{3}{2} \left(-\mathcal{L}_1(u^{n+1}-2u^{n}+u^{n-1}),u^{n+1}-2u^{n}+u^{n-1}\right).
\end{align}
Dropping the last three non-negative terms in \eqref{2.11}, yields \eqref{EnergyProperty-SAV}.

\section*{\large Appendix B. Heuristic analysis of time stepping accuracy}

We adapt the analysis of time stepping in the thin-interface regime $(0 < \varepsilon \ll 1)$ from \cite{CLPW-21}. We consider a general form of the Allen Cahn system
\begin{equation}
 U_t = \mL U +f(U),
\end{equation}
with $\mL = \Delta$ and $f(U)=-\frac{1}{\varepsilon^2} (U^3-U).$
The $\varepsilon$ scaling sets the late-state motion by mean curvature to have normal velocity $V=\mathcal{O}(1)$.  To take a time step from $U^0$ at $t=0$ to $U$ at $t=k$ we have the exact relation
\begin{equation}
\label{e:exactAC}
U=e^{k\mL}U^0 +\int_0^k e^{(k-s)\mL}f(U(s))\,\md s.
\end{equation}
Replacing $f(U(s))$ with $f(U^0)$ yields a first-order scheme ETDRK1. Carrying out the integral
$$ U^*=e^{k\mL}U^0 + \mL^{-1}(e^{k\mL}-I)f(U^0).$$
The error induced by the replacement satisfies 
$$e_0:= f(U(s))-f(U^0)=\mathcal{O}(f_U k U_t).$$
Generically for a quasi-steady front $\Gamma$ the solution $u$ takes the form $U=g(z)$ where $g$ is a front profile and $z=z(x)$ is the $\varepsilon$-scaled signed distance of $x$ to $\Gamma.$  Near the front $U$ is not close to $\pm1 $ and consequently $f_U= \mathcal{O}(\varepsilon^{-2})$. Also near the front $U_t=g'(z)V/\varepsilon=\mathcal{O}(\varepsilon^{-1})$ and $\Delta U = g''(z) (V/\varepsilon)^2=\mathcal{O}(\varepsilon^{-2}).$ The local truncation error has the general size
$$ E_0:=\int_0^k e_0\,\md s= \mathcal{O}\left(k^2/\varepsilon^3\right).$$
This is the same order as that found for the first order IMEX schemes.

Returning to the exact relation \eqref{e:exactAC} the second order replacement
$$ f(U(s))\approx f(U^0)+\frac{s}{k} \left(f(U^*)-f(U^0)\right),$$
yields the second-order ETDRK2 scheme
\begin{equation}
\begin{aligned}
U&= U^* +\frac{1}{k}\int_0^k s e^{(k-s)\mL}\left(f(U^*)-f(U^0)\right)\,\md s,\\
 &=  U^* +\frac{\mL^{-2}}{k} \left(e^{k\mL}-(I+k\mL)\right)\left(f(U^*)-f(U^0)\right).
 \end{aligned}
\end{equation}
The error sources arise first from that implicit in $U^*$ which is $\mathcal{O}(k^2/\varepsilon^3).$ Defining 
$ e_1$ to be the error implicit in $f(U^*)$ we have
$$ e_1=\mathcal{O}\left(k^2/\varepsilon^5\right).$$
Other sources of error in this scheme are smaller, generally $\mathcal{O}(k^2/\varepsilon^4)$.
The associated local truncation error 
$$ E_1:= \frac{1}{k}\int_0^k se^{(k-s)\mL} e_1\md s =
\frac{\mL^{-2}}{k} \left(e^{k\mL}-(I+k\mL)\right)e_1 \approx \frac{k}{2} e_1 = \mathcal{O} \left(k^3/\varepsilon^5\right).$$
 A similar analysis applied to the second-order SBDF2 IMEX scheme shows that the time-stepping error error scales like $\mathcal{O}(k^3/\varepsilon^4).$ 
For a prescribed local truncation error $\sigma$ the time step scaling  for an SBDF2 IMEX code applied to the Allen Cahn system is $k\sim\varepsilon^{\frac43}\sigma^{\frac13},$ (see Table 1 of \cite{CLPW-21}). 
For ETDRK2 the time step scaling on the same system are limited to $k\sim \varepsilon^{\frac53}\sigma^{\frac13}.$ Moreover this work suggests that the scaling gap in the thin-interface error will grow larger for higher order equations such as Cahn Hilliard and the FCH.

%%%% Bibliography  %%%%%%%%%%

\end{document}